# THE NAVAL BATTLE OF ACTIUM AND THE MYTH OF THE SHIP-HOLDER: THE EFFECT OF BATHYMETRY

**Johan Fourdrinoy, Clément Caplier, Yann Devaux** and **Germain Rousseaux,** CNRS – Université de Poitiers – ISAE-ENSMA - Institut Pprime, France
**Areti Gianni** and **Ierotheos Zacharias,** University of Patras, Greece
**Isabelle Jouteur,** Université de Poitiers, Forellis France
**Paul Martin,** Université de Montpellier, France
**Julien Dambrine, Madalina Petcu** and **Morgan Pierre,** Université de Poitiers, Laboratoire de Mathématiques et Applications, France.

**SUMMARY**

A myth of antiquity is explained with modern science in the context of an ancient naval battle. A legend was invoked by the admiral Pliny the Elder to explain the defeat of Antony and Cleopatra against Octavian at the naval battle of Actium. A fish, called echeneis or remora, is said to have the power to stop ships or to delay their motion by adhering to the hull. Naturalists have since studied how the fish sucking-disk with its typical pattern of parallel striae sticks to its host. Here we show the pattern of the free surface measured in a towing tank in the wake of an ancient galley is similar to the striae pattern of the fish. We have measured the bathymetry at the mouth of the Ambracian Gulf that influenced the physical environment of the battle. The computations demonstrate the increase of wave resistance of a galley as a function of the draft to the water depth ratio in shallow water corresponding to the appearance of a particular wake pattern: the echeneidian free surface pattern.

**NOMENCLATURE** [SUMMARY]

| | |
|---|---|
| $\alpha$ | Wake angle (°) |
| $A_n = T/h$ | Antonian number |
| $B$ | Beam (m) |
| $C_F$ | Friction coefficient (N) |
| $\mathrm{Fr}_h = U/\sqrt{gh}$ | Froude depth number |
| $\mathrm{Fr}_L = U/\sqrt{gL}$ | Length Froude number |
| $g$ | Gravity of Earth (m/s²) |
| $h$ | Depth (m) |
| $k = \sqrt{k_x^2 + k_y^2}$ | Wavelength (m⁻¹) |
| $k_x$ | Longitudinal wavelength (m⁻¹) |
| $k_y$ | Transversal wavelength (m⁻¹) |
| $L$ | Length of ship (m) |
| $\lambda$ | Linear scale of ship model |
| $\omega$ | Pulse wave (rad/s) |
| $m$ | Blockage parameter |
| $\nu$ | kinematic viscosity (m²/s) |
| $R_t$ | Total resistance (N) |
| $R_v$ | Viscosity resistance (N) |
| $R_w$ | Wave making resistance (N) |
| $\mathrm{Re} = VL/\nu$ | Reynolds number |
| $\rho$ | Density of water (kg/m³) |
| $S$ | Wetted surface (m²) |
| $\sigma$ | Surface tension (kg/s²) |
| $T$ | Draft (m) |
| $t_W$ | Water temperature (°C) |
| $V_m$ | Model speed (m/s) |
| $V_R$ | Real-scale speed (knots) |
| $UKC_m$ | Under keel clearance |
| $W$ | Tank width |

## 1    INTRODUCTION

September 2, 31 BC was a turning point for the ancient world, and an enigma for historians and scientists of all times. That day, the confrontation of Antony and Cleopatra against Octavian took place, near Actium in the Ambracian Gulf, the epilogue of the Civil War between the Western Roman world and the Eastern Oriental world. Antony, with his heavy fleet composed among others of decaremes, faces Octavian and his light fleet composed among others of triremes. Two mysterious anomalies disrupt the unfolding of history: Antony remains inexplicably motionless for three hours at the exit of the Gulf, then, instead of charging forward to break through the opposing lines, he fails to pick up speed. This forced him to adopt combat tactics involving getting close to the enemy in order to board their vessels for which his large boats were ill-suited. The ancient sources which mention these anomalies either give no explanation at all or give explanations which are less than convincing. Everything seems to point to Antony's fleet first having been compelled to remain motionless and then to Antony's having had to choose the least promising combat tactics. The explanation usually given by historians and modern philologists is that Antony expected wind to rise from the land; then, his fleet, having repelled that of Octavian, could sail off covering Cleopatra who stayed in the rear (Antony's rear fleet was a priority, because it was carrying the war booty). In contrast to the use of naval combat, the fleet of Antony had left the masts and sails lying on the deck of the ship - which has not been easy manned on board, but would allow, when the time comes, to prepare masts and sails to escape and to be sure of not being caught by the enemy ships, equipped to fight, that is to say, only with oars. Things did not go as planned: the collapse of Antony's frontline helped the admiral of Octavian, Agrippa, to attack the isolated part of Antony's vessels. In Section 9.1 with supplementary information on Ancient History, the reasons for the long immobility of the fleet of Antony are examined. We have a clue that this immobility was not expected: the surprise of the opponents. As

Octavian certainly knew the plan of Antony and the tactics he would adopt. He is surprised to see him sitting still. Another clue, less pronounced, is Antony's customary haranguing of his board troops and crews by moving along the front of the ship, but he did this on a small boat, and not - as one would have expected - from his flagship. The purpose was indeed to harangue the troops from a dominant position. In a small boat however, the leader is not in a dominant position… (Carter, 1970; Martin, 1995; Lange, 2011; Murray 2012). The explanation of these events are much written about, and (Tarn, 1931) warned us "The true history of Antony and Cleopatra will probably never be known; it is buried too deep beneath the version of the victors". A legend was invoked by Pliny the Elder (Pliny the Elder, 1857) (the naturalist and the admiral of the western Roman navy in the first century) to explain the defeat of Antony and Cleopatra against Octavian. A fish, called echeneis or remora and ship-holder or sucking-fish nowadays, is said to have the power to stop ships or to delay their motion by adhering to the hull (Jouteur, 2009). Some scientists have brought other reasoning and arguments: biofouling; rudder effect turbulent brake; dead-water in deep water… Gudger (Gudger, 1918) even concluded his review of these explanations with the definite statement that "another myth of the ancients is dissipated in thin air".

A research project has been set up to defy this point of view by analysing three new scientific reasons for the difficulties in manoeuvring by analysing the effect of shallow water only, stratification in shallow water and ship squat. In this introducing work on a scientific study of the battle of Actium, this paper focuses exclusively on the first effect. The bathymetry at the mouth of the Ambracian Gulf that influenced the physical environment of the battle has been measured and is described in Section 2. The Section 3 of the papers contains mathematical computations that demonstrate the increase of wave resistance of a galley with a draft of the order of the water depth in shallow water corresponding to the appearance of a so-called "echeneis" free surface pattern.

## 2 BATTLE'S CONDITIONS
### 2.1 THE OCEANOGRAPHIC CHARACTERISTICS OF THE AMBRACIAN GULF

"Oceanographic research in the Amvrakikos Gulf in Western Greece, a semi-enclosed embayment isolated from the Ionian Sea by a narrow, shallow sill, has shown that it is characterized by a fjord-like oceanographic regime" (Ferentinos et al, 2010). The entrance of the Ambracian Gulf, i.e. the area where the Actium Battle took place, limits the gulf's communication with the open Ionian Sea. It is a particularly shallow and narrow area (see Section 9.4 with supplementary information on oceanography).

The bathymetric map (Hellenic …, 1982) was used to reconstruct the bathymetry of the area where the Actium battle was held. The map's data were digitized and projected in the WGS 1984 - UTM 34N coordinate system. The map's data combined with bathymetric data, that where recovered during two sampling cruises in September – October 2012 period. Depth measurements were made along 10 transects and 20 points, uniformly distributed in the area of interest.

In order to reconstruct the bathymetry (explained in Section 9.4 with supplementary information on Oceanography) at the gulf's entrance in 31 BC, when the battle occurred, basic modifications to the current map were made. These modifications were based on: a) the relative sea level changes during the last 2000 years and; b) the morphological changes due to human interventions in the area over the last decades. The dredging of a navigational channel during the 1970s, changed the area's bathymetry as well as its hydrodynamics and its sedimentation processes. These changes resulted in morphological structures formation, which were identified and removed during the bathymetry reconstruction. In addition, based on literature and observations data, it was concluded that the average sea level during the battle was 75 cm lower than the current one (Lambeck and Purcell, 2005). This was also considered for the ancient bathymetry reconstruction. Decaremes (namely the biggest boats at Actium, see below), with a draft of 2.1 m, have been limited in their position. Indeed, part of the entrance has a depth of less than 2.5 m.

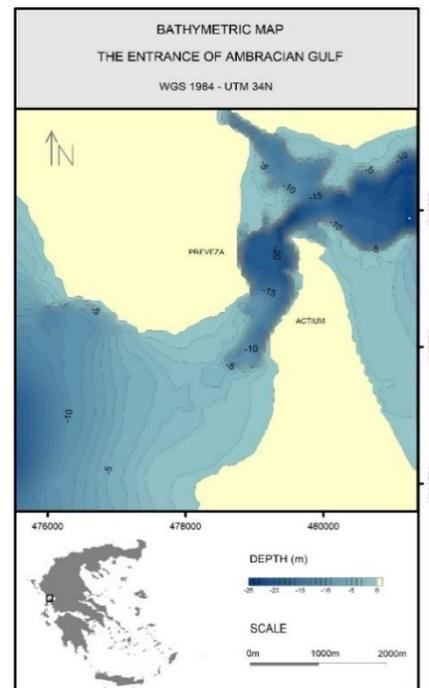

**Figure 1. The bathymetric map of the Ambracian Gulf entrance 2000BP.**

### 2.2 RECONSTRUCTION OF AN ANCIENT GALLEY MODEL

Laboratory experiments have been carried out in order to reproduce the assumed configuration of the battle. The water level, vessel's speeds and dimensions have been determined using the aforementioned bathymetry measurements, naval archeo-architecture inputs and historical reports of the battle (see Section 9.1 with supplementary information on ancient history and Section 9.3 with supplementary information on naval architecture). Both fleets in presence at Actium had very different characteristics with respect to naval architecture. Hence, we decided to take as representative classes of boats for the two fleets: a trireme for Octavian and a decareme for Antony, both featuring the Athlit ram. According to (Murray et al., 2017), the Athlit ram would belong to a class 4. However, because of the reduced size of our warship model in the experiments, differences in the ram's geometry and epoch would be negligible at these scales.

It seems there is a consensus around the naval plans of a trireme with a slight variation depending on the period: the fifth-century BC trireme Olympias has dimensions a little bit smaller than the triremes present during the first century BC Actium battle. Unfortunately, there is no historical evidence for the real dimensions of a decareme. As a matter of fact, the boat classes bigger than 5 were no more built after the battle of Actium, principally because of Antony's defeat.

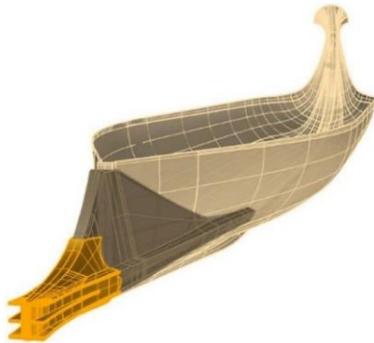

**Figure 2. The small scale model of the Greek galley used in the experiments.**

The geometry of the reduced model was based on the hull lines of the trireme Olympias generously provided by the Trireme Trust (Figure 2). At the water line, the ship model is 120 cm long and 13.5 cm wide. The draft is around 3,9 cm, depended of configurations (trireme or decareme), and the T/h ratio. The ram geometry was reverse engineered by (Murray, 2012) and the Institute for Visualization of History. As we use the same model to study the trireme and decareme behaviours, we scale the experiments using scaling laws (Table 1). The Olympias trireme is 32.08 m long, 3.43 m wide, with a 1.05m draft at waterline, and as presented in the main text we suppose that decaremes were twice as big, so the respective scaling-factor for the lengths is 26.73 for the trireme configuration and 53.47 for the decareme one. Considering the bathymetric data presented in the supplementary information on Oceanography, we chose 3 meters as a representative mean water depth at the outlet of the Ambracian Gulf, when corrected for the change in the water level since the battle. Hence, the water height in the towing tank was set to 11.22 cm for the trireme configuration and 5.61 cm for the decareme configuration (so the underkeel clearance is respectively 7.27 cm and 1.68 cm). The model speeds have been Froude depth numberscaled with (water depth) Froude number. Experiments were carried out for the height and length Froude numbers values indicated in the table below. The speed values are given in meter per second for the model and in knots for a real-scale vessel. During the experiments, the water temperature was about 21°C. Calm water resistance tests with a small-scale trireme model had already been carried out in the past by Grekoussis and Loukakis (Grekoussis and Loukakis, 1985, 1986) with a 3.2 m long small-scale model in a water depth of 3 m. The range of the Froude numbers $Fr_L$ was between 0.090 and 0.397, corresponding to Froude depth numbers $Fr_h$ between 0.093 and 0.410. Given these values we can assert that their experiments were performed in deep water conditions and they did not focus on shallow water effects. The choice of the 1.2m length for our small-scale model allows us to explore a wider range of Froude depth numbers (between 0.3 and 1.63), while staying under the limit length Froude number $Fr_L=0.5$ recently highlighted by (Rabaud and Moisy, 2013; Noblesse et al. 2014) from which the angle of the wake starts to decrease (an effect already present in deep water). Hence the maximum of wave resistance measured corresponding to the appearance of a shallow water wake pattern and not to another phenomenon. In battle conditions, if boats were in compact formation, a lateral confinement effect (as in tank) can be envisaged. There would be interference between the wakes, which will be the subject of a future. The limitations of the reduced model were studied in section 9.6.

**Table 1. Significant values for trireme and decareme with reduced and real scales. MS=model scale / RS=real scale.**

|  | Trireme RS | Trireme MS | Decareme RS | Decareme MS |
|---|---|---|---|---|
| $\lambda$ (ratio scale) | 26.73 | | 53.47 | |
| $L$ (length) | 32.08 m | 1.2 m | 64.15 m | 1.2 m |
| $B$ (beam) | 3.6 m | 13.5 cm | 7.2 m | 13.5 cm |
| $T$ (draft at midship) | 1.05 m | 3.93 cm | 2.10 m | 3.93 cm |
| $h$ (depth) | 3 m | 11.2 cm | 3 m | 5.6 cm |
| $UKC$ (under keel clearance) | 1.95 m | 7.2 cm | 0.9 m | 1.68 cm |
| $A_n = T/h$ | 0.35 | 0.35 | 0.70 | 0.70 |
| $W$ (tank width) | Min: 5×B=18 m (compact formation) | 1.49 m | Min: 5×B=36 m (compact formation) | 1.49 m |
| $m$ (blockage parameter) | 0 | 0.013 | 0 | 0.026 |
| $V$ (boat speed) | 5 knots | 0.50 m.$s^{-1}$ | 5 knots | 0.35 m.$s^{-1}$ |
| $V$ (boat speed) | 10 knots | 1.00 m.$s^{-1}$ | 10 knots | 0.70 m.$s^{-1}$ |

## 3 PHYSICAL COMPARISON OF TRIREME AND DECAREME CONFIGURATIONS

### 3.1 MATHEMATICAL COMPUTATIONS OF THE WAVE RESISTANCE OF ANCIENT GALLEYS

From naval architectural data and based on Sretensky's analytical formulation (Sretensky, 1936), it is possible to calculate a prediction of wave making resistance of an ancient galley based on linear theory. Because of the importance of the ships design in the battle of Actium, we took into account the actual shapes of the galleys. Until now, numerical computations of the wave resistance with Sretensky's formula involving real ship hulls were made by using polynomial representation or uniform grids. In our case, the ships exhibit details at different scales. This led us to use meshes with triangular elements, refined in areas of finer details such as the ram at the bow of the ship (see Section 9.5 with supplementary on mathematics). The theoretical predictions of Sretensky require to be in shallow water configuration, without significant hydraulic effects (water level drawdown and return current) (Pompée, 2015). The numerical calculations present in Figure 3, were carried out for trireme or decareme configurations, and by varying the Antonian number $A_n=T/h$. As observed by (Russell, 1839; Inui, 1954), we observe a peak of resistance for $Fr_h = 1$ whose magnitude grows with $A_n$. To this resistance we can add a viscous resistance due to the friction of the boat with the water. This viscous resistance can be predicted by the (ITTC, 1957) protocol. By adding these two components of resistance (wave and friction), we obtain a total resistance according to the speed, the geometry (decareme or trireme), and the Antonian number (Figure 3). See SI on Mathematics for curves showing viscous and wave contributions for each configuration.

Using the measured bathymetry and the previous computations, and last results, we infer the theoretical wave making resistance of both the trireme and decareme in various points of the mouth of the Ambracian Gulf. The results of our predictions on total resistance are summed up in the two maps shown in the Figure 4. Our maps were computed for two velocities: 7 knots (left figure) and 10.5 knots (right figure), so $Fr_h=1$ when $h=3$ m. The latter being typical of a ramming manoeuver whereas the former corresponds to the cruising speed. These maps show in colours $R_D/R_T$, i.e. the ratio between the total resistance applied to a decareme and the one applied to a trireme in each point of the Ambracian Gulf. Our measured bathymetric data are plotted with line contours (in white), and three particular areas are highlighted: the shallow zone inaccessible for the decareme (in grey), the shallow zone inaccessible for both the trireme and the decareme (in black), and the land (in brown). At the cruising speed of 7 knots, our calculations predict a wave resistance ratio close to 1, almost uniformly on the battlefield, which means that no particular ship has an advantage when its velocity is lower than the ramming velocity (see Section 9.5 with supplementary on mathematics). At this speed, the viscous resistance is the main component of the total resistance. Hence, the factor 4 as explained in the SI which is mostly compensated by the ratio in the number of rowers 605/170=3.56. Thus, the larger wetted surface of a decareme is compensated by a greater rowing power. The trap is a confinement effect, not a simple viscous effect. At the ramming speed of 10.5 knots, the wave resistance ratio is much higher and the $R_D/R_T$ can go up to, forming a bottleneck zone at the entry of the gulf. This result confirms the idea that ramming may have been impossible for Antony's ships in the particular entrance zone of the Ambracian Gulf and hence answers to the second anomaly underlined by the historical reports namely the impossibility to use the ramming tactic.

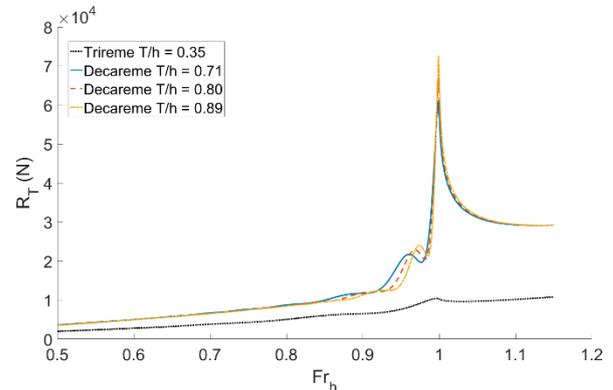

**Figure 3.** Calculated total resistances, composed by a wave making resistance and a viscous resistance, as a function of $Fr_h$ for a varying ship draft to depth ratio.

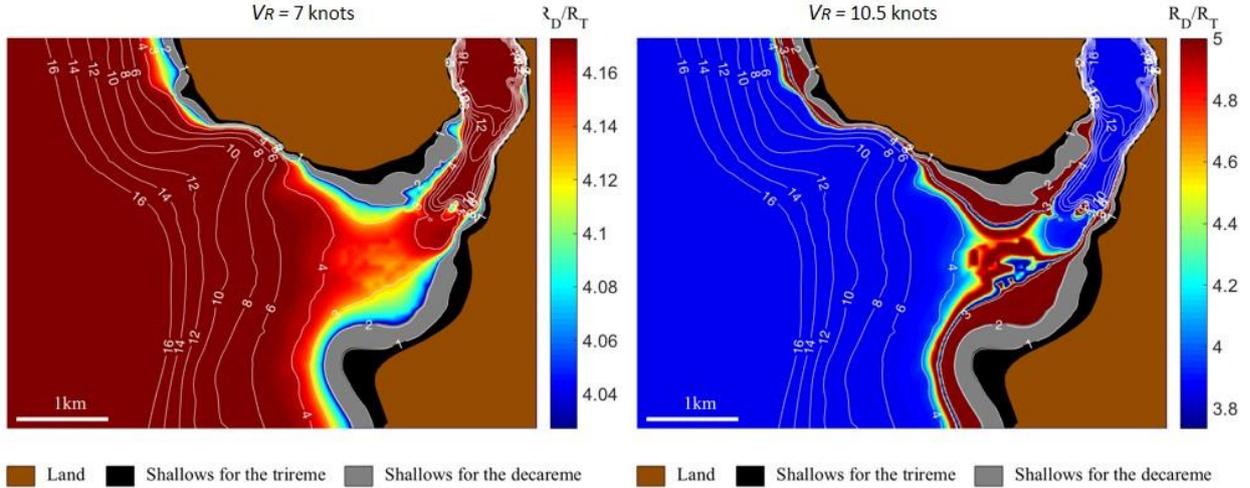

**Figure 4.** Maps featuring the ancient bathymetry and theoretical predictions of total resistance ratio $R_D/R_T$ for two different velocities: 7 knots (left) and 10.5 knots (right). For the attack speed of 10.5 knots, decareme's resistance $R_D$ is two to ten times larger than the trireme's resistance $R_T$. The colormap has been limited to 5 to make the results more visible, however, in a small area at the entrance of the channel, the ratio $R_D/R_T$ may increase up to a factor 10 at 10.5 knots (see SI on Mathematics for an unlimited colormap).

### 3.2 THE SURFACE WAKES OF AN ANCIENT GALLEY

Experiments are carried out in a towing tank 20 m long and 1.49 m wide. The model, placed in the middle of the tank, is fixed to prevent any degree of liberty, and test only the impact of the draft. The top-view of wake is recorded by fast camera at 125 Hz. The wake pattern of the ship gives a clue on the deep or shallow water configuration. Indeed, in deep water, the usual Kelvin wake pattern, featuring a V shape, has a constant angle at 19.47°. From $Fr_L \approx 0.5$, the angle should diminish with the Froude number based on the length of the boat. (Rabaud and Moisy, 2013) propose a decrease according to $1/Fr_L$ while (Noblesse et al., 2014) propose a decrease according to $1/Fr_L^2$. In shallow water, this angle is dependant of the Froude depth number, reaching a maximum for $Fr_h = 1$ (Havelock, 1908; Inui, 1936; Soomere, 2009; Ersan and Beji, 2013). On the top views, one observes two V-like wakes at the bow and the stern of the galley. Each wake is usually composed of a system of divergent and transverse waves, which superimpose and form the so-called cusp waves, defining an envelope corresponding usually to maximum wave heights (Darmon et al., 2013): here, we observe mainly the diverging wave system, as was observed in the trials of Olympias in the nineties (Morrison et al., 2000). In addition, a turbulent wake is clearly seen behind the stern of the ship. There is a measure of the wake angle in the spectral domain, from an image of the wake seen from above (Figure 6 and see 9.6.b for the methodology). The dispersion relation:

$$0 = V_m^2 k_x^2 - \left(gk + \frac{\sigma}{\rho}k^3\right)\tanh(kh)$$

has an inflexion point, where the slope is directly connected to the angle (Carusotto and Rousseaux, 2013; Gomit, et al. 2014; Caplier 2015). The trireme and decareme configurations show the same evolution of the wake angle, corresponding to a shallow water regime (Figure 7). This validates the hypothesis necessary for the use of the Sretensky's formula to be in shallow water in order to compute the wave resistance.

For the decareme, an additional system of quasi-parallel waves of the divergent type appears in addition to the Kelvin wake pattern and superimposes to create another wake pattern starting roughly at $Fr_h = 0.8$ (Figure 8). The amplitude of this additional system of waves increases with respect to the speed of the ship until a Froude depth number of 1.0, where they are the most visible. Past that Froude number, the value of the amplitude decreases with the speed, and the quasi-parallel waves bend toward the stern-wake (Figure 8). The Froude depth number of 0.85 corresponds to the real scale speed around 9-10 knots, which was approximately the attack speed of the galleys. In addition to this "echeneidian" wake-pattern, a double bow wave appears, only in the decareme configuration, with a similar behavior and a maximum amplitude reached for $Fr_h = 1.15$ (Figure 9). The first wave (in green) presents an angle similar to a Mach cone, while the second (in red) does not seem linear. If the boat stops suddenly, the second wave unfolds and is ejected forwards (Figure 10). This behaviour is reminiscent of the shallow wave pattern of the free surface observed a long time ago by the engineer Scott Russell (Russell, 1839) who towed boats in a shallow canal of Scotland. On the contrary, the first wave (in green) remains folded. Thus, the first wave can be another shallow effect (amplified by a horizontal confinement). The second is due to a canal effect, so therefore absent at sea.

The modification of the wake shape is known since then to be related to an enormous increase of wave resistance (Inui, 1954; Kirsch, 1966; Kostyukov, 1968). This typical shallow water wake behind the galley is strikingly similar to the echeneis suction disk (Figures 12, 13) that was reported to have appeared during the Actium battle when interpreting correctly Pliny's and Octavian's accounts, or as described by Elien (Elien, 1972): "For adhering with its teeth to the extreme stern of the ship driven by a following wind and full sails, just as an unmastered and unbridled horse is held in with a strong rein, so the fish overcomes the most violent onset of the winds and holds the ship as if tied fast to her wharf. [...] But the sailors understand and realize what ails the ship; and it is from this action that the fish has acquired its name, for those who have had experience call it the Ship-holder".

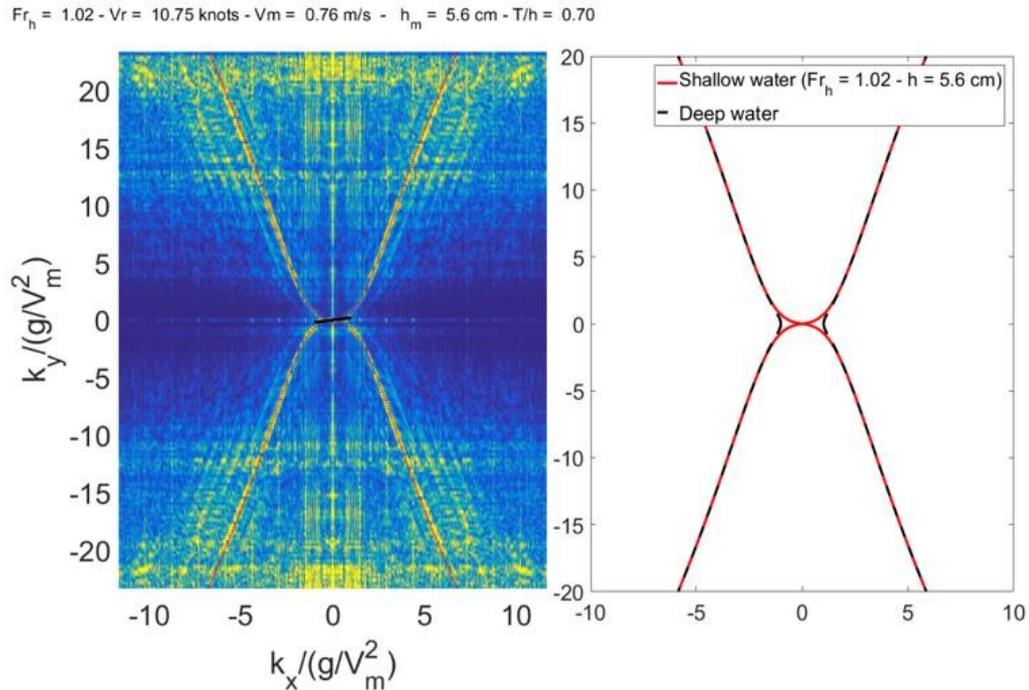

**Figure 6.** Left: Fast Fourier transform of the surface wake based on simple visualization with an aerial picture, in configuration decareme at $Fr_h=1.02$. In red, the theoretical dispersion relation; in black the slope at the inflexion point. Right: Theoretical dispersion relation at $Fr_h=1.02$, in shallow and deep water configurations.

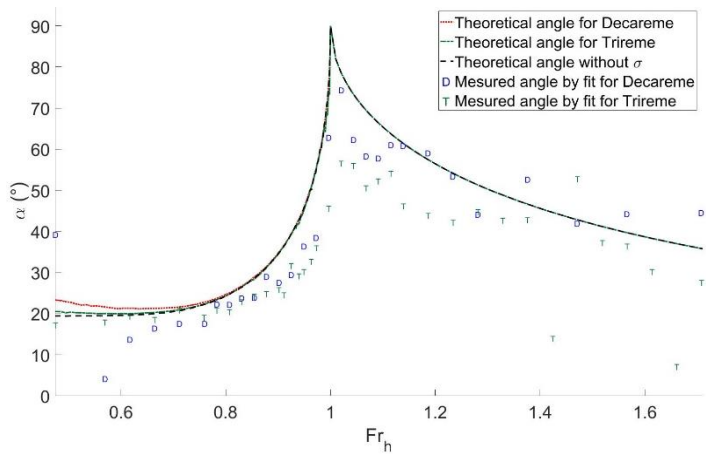

**Figure 7.** Measured wake angle (Kelvin angle) via spatial FFT, in configurations decareme and trireme, as a function of $Fr_h$. Green T are measured in trireme configuration and blue D in decareme configuration. Black curve is the theoretical angle's behavior by (Havelock, 1908), valid for linear theory with an idealized point source.

While the dimensionless number $Fr_h$, which is identical for trireme or decareme, only indicates a shallow water behavior, we use the Antonian number $A_n=T/h$, which drives the effect of strong vertical confinement. By adjusting this number, by increasing the draft, we observe a similar but amplified behavior of the particular wake pattern (Figure 11), as calculated by the formula of Sretensky (Figure 3). It is said that Antony's boats loaded with both sails and war chest (Carter 1970; Martin, 1995; Lange, 2011), which would imply a stronger draft.

In addition to these simple visualizations, the whole wave field behind the boat has been experimentally measured in the towing tank by a stereorefraction method (Caplier, 2015; Gomit, 2013) (Figure 12). This method is based on the calculation of the surface undulations from the apparent deformation of a pattern (roughcast) placed on the bottom of the towing tank. The refraction of light rays at the water-air interface allows, through two cameras, to reconstruct a 3D visualization of the wake. This reconstruction of the free surface deformation due to the motion of the ship clearly highlights the non-classical wake pattern that has been observed and identified behind the decareme in the experiments. The complexity of the wake pattern would have been impossible to capture with classical intrusive local methods such as resistive or acoustic probes so it was necessary to use this state-of-the-art optical method to measure the whole wake (more information in the SI on Fluid Mechanics).

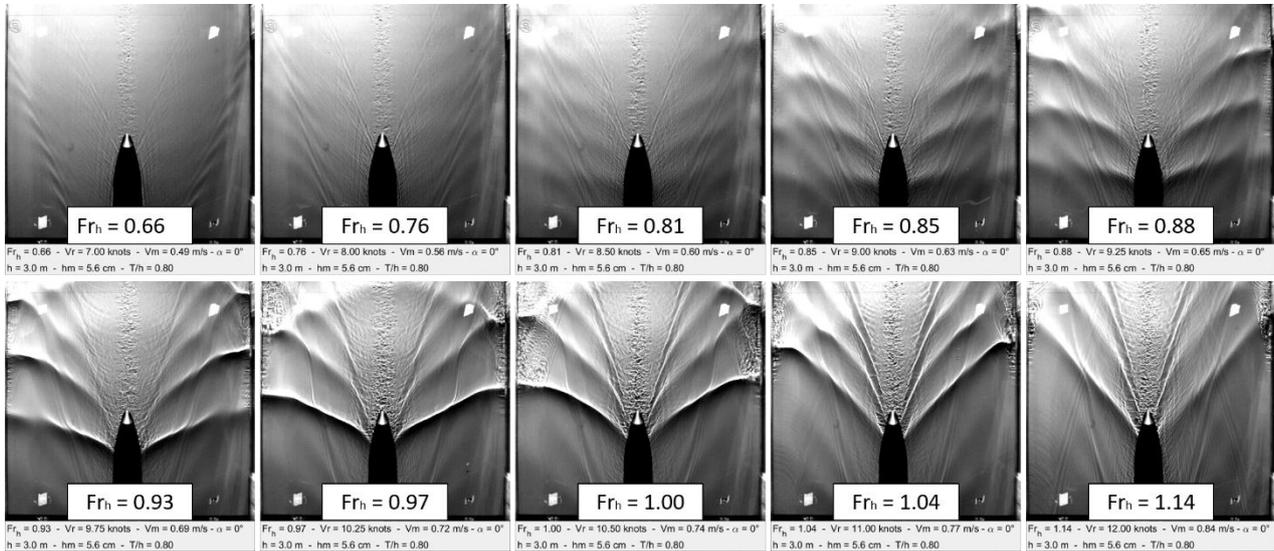

**Figure 8. Evolution with Froude depth number of the wake pattern for a decareme configuration. Supplementary to usual Kelvin wake pattern, quasi-parallel waves of the divergent type appear from $Fr_h > 0.8$. The amplitude of this additional system of waves increases with respect to the boat speed until a $Fr_h = 1$. After, the value of the amplitude decreases with the speed, and this parallel waves bend toward the stern wake. The ship has no angle with the horizontal.**

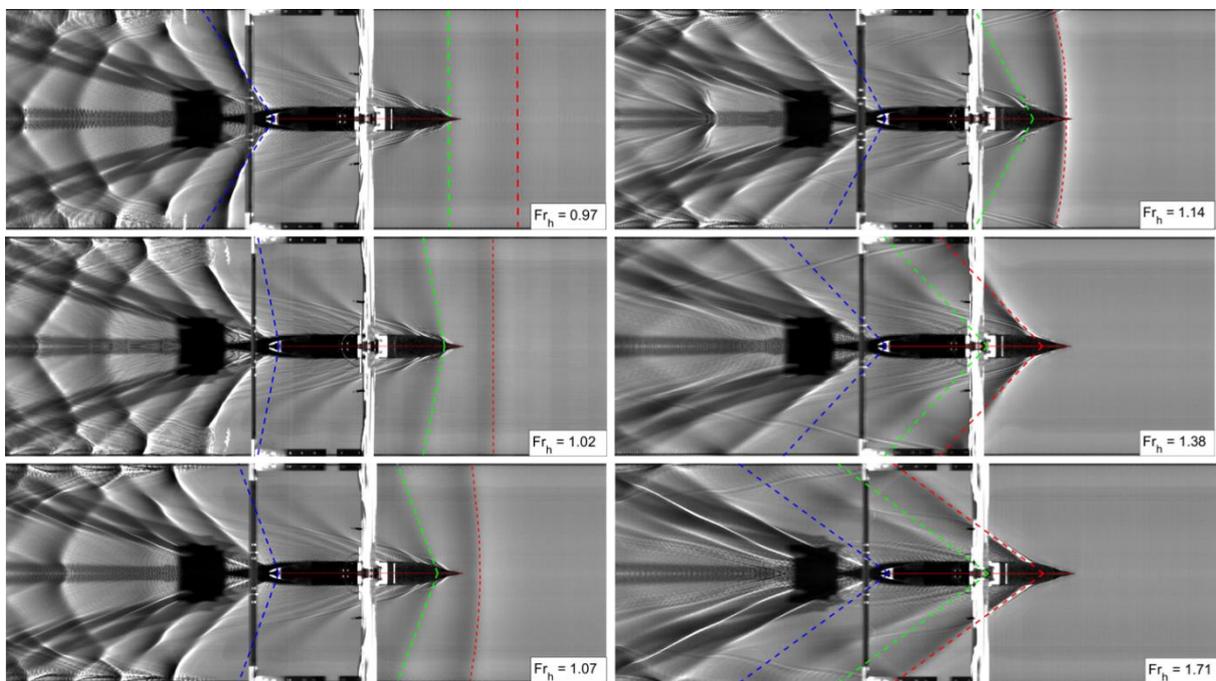

**Figure 9. Evolution with Froude depth number of the bow wave for a decareme configuration. A bow wake (in green) appears from a $Fr_h \approx 1$. Its angle decreases according to a Mach angle : $\alpha = \tan^{-1}\left[Fr_h^2 - 1\right]^{-1} = \sin^{-1}[Fr_h]^{-1}$. A second bow wave with a bigger amplitude (in red) appears in the front of the first.**

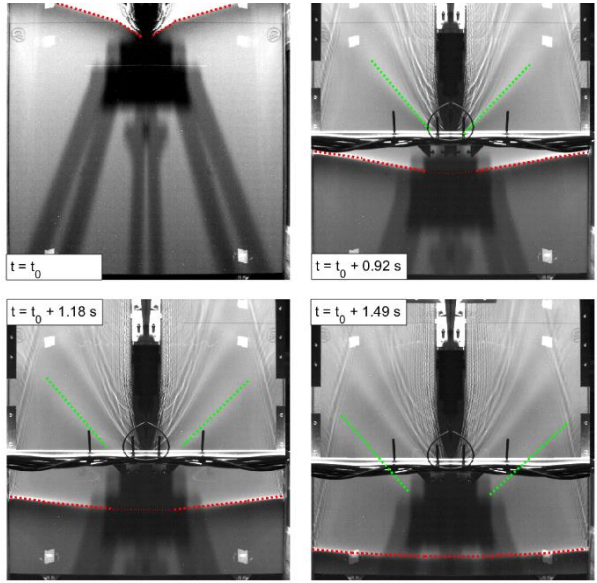

**Figure 10:** Top-views of bow wave ejections when the boat decelerates before stopping at the end of the run. The first bow wave (in green) keeps its slope and is a purely linear shallow effect. The second bow wave (in red) detaches from the prow and creates a solitonic wave à la Scott Russell (Scott Russell, 1939). The second wave is usually akin to a channel effect (not present in the open sea).

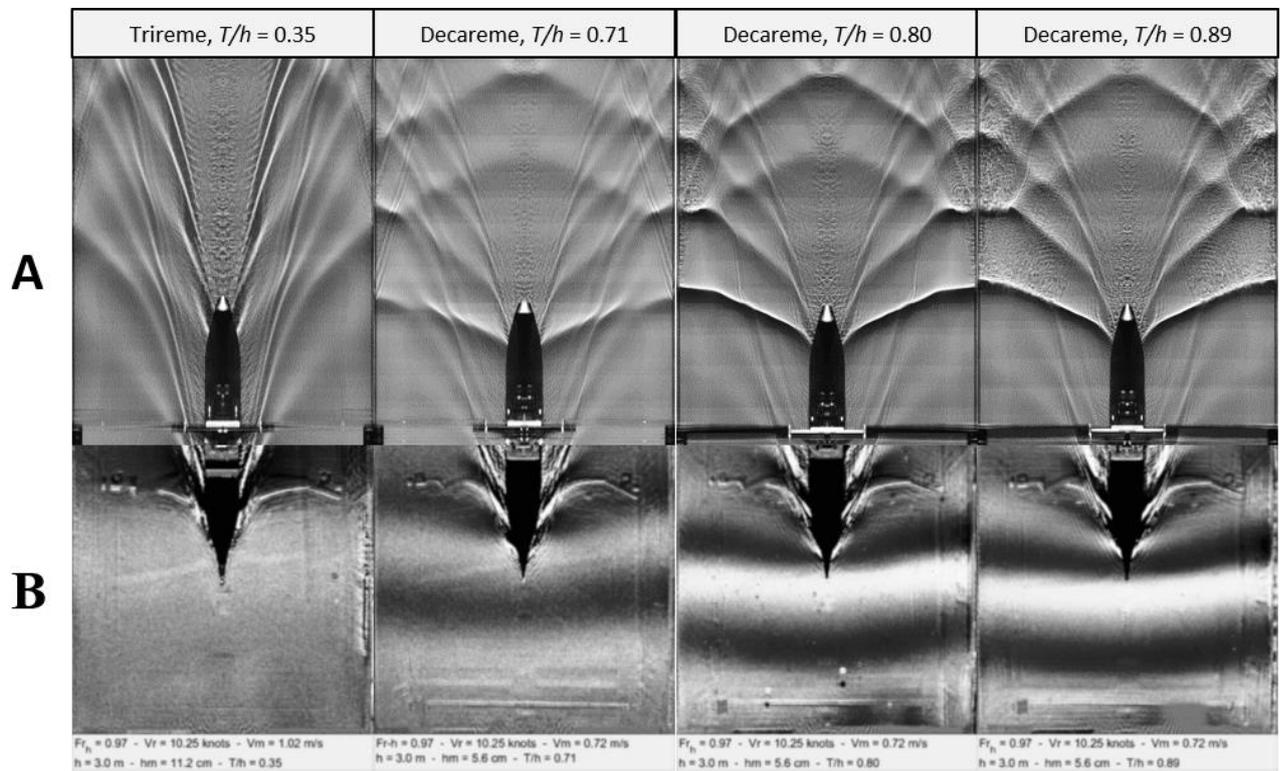

**Figure 11.** Top-views of the wake pattern of the ship for a trireme and a decareme at $Fr_h=0.97$. (A) Unlike the wake pattern of the trireme which is similar to the usual Kelvin wake pattern, in decareme configuration we observe the "echeneidian" pattern (quasi-parallel waves of the divergent type) superimposes with the deep water wakes and which create a complex wake pattern. The amplitude of this shallow waters wake increases with $T/h$. (B) A bow wave appears in decareme configuration (not for a trireme) whose amplitude increases with $T/h$.

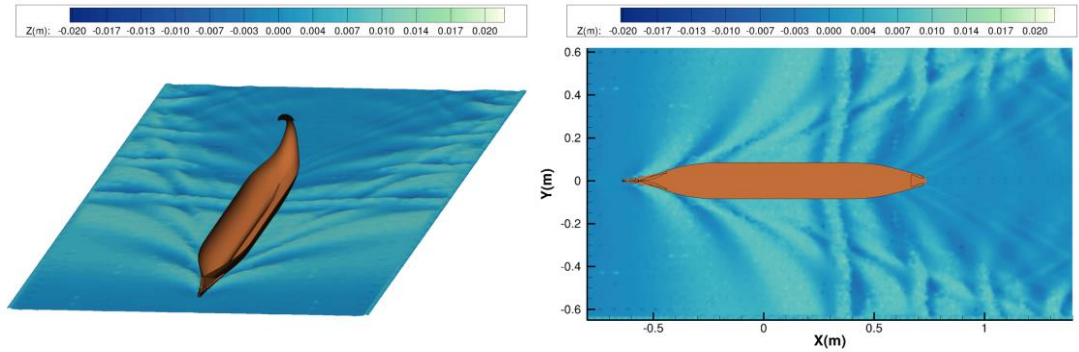

**Figure 12:** The particular echeneidian wake behind the ship in the decareme configuration at Fr$_h$=0.85, measured with the stereorefraction method in the towing tank. In this experiment the keel of the boat makes an angle of 0.13 ° with the horizontal (stern sunk) which is equivalent to an increase of an effective draft along the hull.

## 4   DISCUSSION: THE SCIENTIFIC EXPLANATION OF THE LEGEND OF THE SHIP-HOLDER

According to Albert Günther, "there is scarcely a fish of the existence of which the ancients have been equally certain, and which has so much occupied their imagination... as the Echeneis of the Greeks or Remora of the Latins" (Günther, 1860). With our interdisciplinary approach between human and fundament al sciences, we believe we can explain this famous myth of Antiquity: the battle of Actium, where an echeneis, the small fish which allegedly hampered ships and triggered the interest of historians, writers and poets for twenty centuries. For example, Ovid in his Halieutica, says "The small echeneis is present, wonderful to say, a great hindrance to ships". As we have seen in the introduction, this myth is invoked to explain, e.g., the defeat of Antony and Cleopatra against Octavian at the naval battle of Actium twenty centuries ago. The admiral Pliny the Elder reports: "At the battle of Actium, it is said, a fish of this kind stopped the Pretorian ship of Antony in its course, at the moment that he was hastening from ship to ship to encourage and exhort his men, and so compelled him to leave it and go on board another. Hence it was, that the fleet of Ceasar (Octavian) gained the advantage in the onset, and charged with a redoubled impetuosity" (Pliny the Elder, 1857).

The issue of the origin of the echeneis tale, which is said to have detained Antony, is discussed in Section 9.2 with supplementary information on Linguistics. According to the common opinion, the legend was created by the defenders of Antony and intended to explain the immobility of the flagship, and thus that the fleet did depend on the flagship's moves. For our part, we believe, on the basis of a number of indications contained in the poetic exaltation of the contemporary battle of Actium, that it is one of the themes of Augustan propaganda on this battle, which was exalted as the Principate epiphany. The legend of echeneis is prior to Actium and it was applied to the excitement of the battle, to show that the gods and nature itself were on the "good side", that of Octavian. It is known that Octavian Augustus, after the battle, founded the town of Nicopolis ad Actium; on this forum stood, as in Rome, a forum rostra (one can see the remains today) adorned with the rams of several Antonians ships (Murray, 2012), including probably the flagship of Antony (abandoned by him for a faster one, a quinquereme, at the time of flee). Our hypothesis is that when the ship was lifted from the water in order to recover the rostrum, an echeneis was found attached to the hull and this served to support the activation of the legend. The fish is known to stick to rock or boat in bad weather and the Actium battle happened after four days of storms. Plutarch gives details on the progress of the naval battle: "Caesar (Octavian)… was astonished to see the enemy lying motionless in the narrows; indeed, their ships had the appearance of riding at anchor" (Plutarch, 1988). One of the possible interpretation of the use of the word anchor relies on the legend of the "echeneis" from *echein*- (to hold) and -*naus* (the ship). Pliny the Elder, who was a naturalist and natural philosopher as is well known nowadays but also the Admiral of the fleet of Mycene in the Mediterranean Sea which is less known, gave an explanation for the difficulties that the galleys of Antony had to struggle with during the naval confrontation by invoking the Greek myth. The fish, called echeneis in Greek or remora in Latin is said to have the power to stop ships or to delay ("mora" in Latin) their motion by adhering to its stern. Naturalists have since studied the way in which the haustellum (a sucking-disk with a typical pattern of parallel striae) of the fish exerts an enormous pressure on its host (sharks, turtles, whales, boats, scuba divers, etc.). For Pliny's translator, J. Bostok, the echeneidian myth is "an absurd tradition, no doubt, invented, probably to palliate the disgrace of defeat". But for others, the authority of Pliny the Elder (and subsequent commentators) is such that he would not have relied on such a tale even to protect the reputation of Antony and Cleopatra.

The purpose of this work has been to provide a visual explanation of the legend (Figures 12, 13) which corresponds to the naval difficulties met by the Antonian fleet based on scientific clues. Hence, we have displayed for the first time the visual signature of the Ancient myth of the echeneis, which answers the first anomaly noticed by the historians in the introduction and substantiates the legend for the linguists.

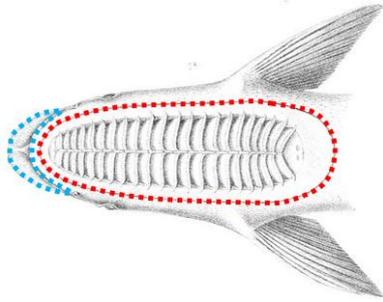

**Figure 13.** Illustration of Echeneis naucrates (Grandidier, 1885). The lips (blue) can represent the bow waves, and the sucking disk pattern (red) can represent the particular echeneidian wake pattern.

## 5 CONCLUSIONS

For the first time since twenty centuries, we have shown conclusively that the global pattern of the free surface measured in this work with modern and non-intrusive optical methods in the wake of an ancient galley moving in shallow waters is similar to the pattern of striae on the sucking-disk of the echeneis fish. Hence, the Antonian boats have been influenced by a physical echeneis and not a biological one during the battle of Actium. From the analysis of the resistance charts, we have demonstrated that the Antonian fleet was unable to use the ramming tactics because the wave resistance was increased up to ten times compared to the Octavian fleet. By a strange coincidence (or maybe not a hazard?), several centuries later another naval battle at the same location produces the same astonishment for the final result: Preveza battle in 1538, where the Ottoman forces fought against the Christian navy and, to the general surprise, won. The Ottoman fleet under the command of Barbarossa with the smallest boats albeit considered as inferior, prevailed. Another possible explanation for the boats difficulty in manoeuvres is the dead-water phenomenon, which can be encountered, for example, in the Northern fjords where ice melting creates two water layers of different densities, with a sharp interface between fresh and saline water (see Section 9.4 with supplementary on oceanography and Ekman, 1904; Grue, 2015; 2016; Esmaeilpour, 2017). The resulting wave resistance exerted on moving boats is significantly increased by the generation of internal waves at the interface. Hence, our future goal will be to compute theoretically the wave resistance in a two-layer shallow water basin since the mouth of the Ambracian gulf has a shallow fjord recirculation. In the laboratory experiments, we will measure the wave resistance of ship models (corresponding to the Actium and Preveza battle) moving in our towing tank at different density stratifications. In addition, confinement effects like water level drawdown, return current, ship squat will be examined as well. Thus, we hope to shed light on the History of these two naval battles with the help of Contemporary Science… With respect to the biological remora, it is unable to explain, of course, the drag on the boat (see Beckert et al., 2016 for a recent study on the fluid dynamics of an attached remora) … We anticipate our work will allow a revisit by historians of the events and by linguists of the legend as well as open new perspectives on battles with similar conditions like the one of Preveza in 1538.

## 6 ACKNOWLEDGEMENTS

This research was funded by the CNRS Interdisciplinary Mission in 2013 (INSIS/INSHS co-funding), by the Poitiers University through two successive two-years funding (ACI UP on Wave Resistance 2013-2014 and ACI Pprime on Actium 2014-2015) and by the Patras University (Inland & Coastal Water Laboratory). The authors are indebted to William Murray and the Institute for Visualization of History for sharing their 3D plans of the Athlit rostrum and those of the temple in Nicopolis. The hull lines of the galley were provided by the Trireme Trust and we thank Doug Pattison for discussions on naval architecture. Christian Oddon from Cabinet Mauric translates the drawn hull lines of John Coates into numerical lines with modern softwares of naval architecture. The reduced model was built by Formes & Volumes in La Rochelle. We thank the services of metrology, computer sciences, electronics and the mechanical workshop of the Pprime Institute in Poitiers for helping in the course of experiments. The support of Michel Briand is gratefully acknowledged. We thank the former and current directors of the Pprime Institute and of the LMA. This work benefited from the support of the project OFHYS of the CNRS 80 Pprime initiative in 2019.

# 8 AUTHORS BIOGRAPHY

**Johan Fourdrinoy** holds the current position of PhD student at University of Poitiers - Institut Pprime, France. His thesis is about wave-current interactions. His previous experience includes engineering fluid mechanics and the study of a wave maker.

**Clément Caplier** holds the current position of post-doctoral researcher at the University of Poitiers - Institut Pprime, France specialized in optical measurement techniques for hydrodynamics. His previous experience includes an experimental thesis on the effects of finite water depth, lateral confinement and current on ships wakes and drag.

**Yann Devaux** holds the current position of teaching and research assistance at University of Poitiers - Institut Pprime, France. His thesis is about the unsteady suspension of sediment. His previous experience includes the establishment of an experimental process for studying dead-waters.

**Areti Gianni** is geologist – marine science researcher focusing on degraded ecosystems management and restoration. She is a technical consultant in OIKOM - Environmental Studies LTD and a postdoctoral researcher at the Patras University. Her research is focused in inland and coastal ecosystems' hydrodynamics, biogeochemistry, eutrophication and anoxia. She is specialized in hydrodynamic and ecological modelling in inland and coastal waters, and in the development of management/restoration practices and techniques.

**Ierotheos Zacharias** is an oceanographer focusing on marine ecosystem monitoring and restoration. He is an Associate Professor in the University of Patras, Greece. His research is focused in inland and coastal ecosystems' hydrodynamics, biogeochemistry, eutrophication and anoxia. He is specialized in hydrodynamic and ecological modeling in inland and coastal waters, and in the development of management/restoration practices and techniques.

**Isabelle Jouteur** holds the current position of Professor of Classics at University of Poitiers (Forellis). Her works focuses on Latin poetry of the Augustan period; she is interested in mythology and imagination with an anthropological point of view. Her previous experience includes the study of monsters and wonders in Greco-Roman culture.

**Paul Martin** is Professor Emeritus of the University of Montpellier-III and Honorary President of the magazine Vita Latina. Specialist in the history of royal and republican Rome, he was particularly interested in the process of return to the monarchy that led to the imperial regime. As such, he had to deal with the founding battle of the August regime: the battle of Actium, including the strange phenomenon reported by several ancient sources: the immobilization of the ships of the fleet of Antony at beginning of this battle.

**Julien Dambrine** holds the current position of Assistant Professor at the University of Poitiers. He is responsible for the prediction of the wave-making resistance shown in this article using actual antique galley hull shapes. His speciality is shape optimisation in the context of ship waves, which include the aforementioned predictions.

**Madalina Petcu** holds the current position of Associate Professor (maître de conférences, HDR) at University of Poitiers, Laboratoire de Mathématiques et ses Applications. She is member of the Partial Differential Equations team, working on the theoretical and numerical aspects of mathematical models in fluid dynamics and other models related to physics. She has also a full-time teaching duty, consisting of 192 hours of teaching for the Mathematics Department of Poitiers University.

**Morgan Pierre** holds the current position of associate professor at University of Poitiers (France). He is responsible of the Partial Differential Equations team at the Laboratoire de Mathématiques Appliques (Department of Applied Mathematics). His previous experience includes work on optimisation of ship hulls and wave resistance.

**Germain Rousseaux** is a senior research scientist at CNRS – Institut Pprime, France. He is a proponent of interdisciplinary studies like for example his recent works on analogue black holes with the measurement of Hawking radiation in the laboratory.

# 9 SUPPLEMENTARY INFORMATION
## 9.1 SUPPLEMENTARY INFORMATION ON ANCIENT HISTORY

What follows is based on a critical examination of ancient sources relating the battle of Actium between Antony's and Octavian's fleets on 2nd September 31 BC at the entrance to the gulf of Ambracia. The main sources are Plutarch's *Life of Antony,* 61- 68 and Dio Cassius's *Roman History,* 50, 14-35. Plutarch mentions one particular fact, which greatly puzzled those on Octavian's side. This happened before the battle when the two fleets were facing each other. Octavian and his admiral Agrippa were surprised to note that Antony's fleet remained at a standstill for at least three hours, until midday, instead of, as was customary in ancient times, attacking at dawn:

« *Caesar (Octavian) [...] was* **astonished to see the enemy lying motionless in the narrows; indeed, their ships had the appearance of riding at anchor. For a long time he was convinced that it was really the case,** *and kept his own ships at a distance of about 8 furlongs from the enemy. But it was now* **the sixth hour, and since a wind was rising from the sea,** *the soldiers of Antony became* **impatient of the delay,** *and, relying on the height and size of their own ships as making them unassailable, they put their left wing in motion. When Caesar saw this* **he was delighted,** *and ordered the right wing to row backwards, wishing to draw the enemy still farther out from the gulf and the narrows, and then to surround them with his own agile vessels and come to close quarters with ships which,* **owing to their great size and the smallness of their crews, were slow and ineffective** (1). »

In fact, it was more the crews' lack of experience which could be a real handicap and it did, indeed, hinder Antony's fleet during the battle as his ships had to move and turn in the midst of enemy vessels. On the contrary, the maneuver necessary for a forward attack using the *rostra* was well within the capability of inexperienced crews: all they had to do was to launch the ships at full speed straight ahead, towards the enemy -a good enough reason to opt for this tactic and yet Antony did not choose to do so. Plutarch suggests a possible explanation for this strange immobility: Antony would have been waiting for the wind to get up so as to take advantage of it to escape. Modern historians have religiously followed his lead in this matter and given this out as the reason for Antony's decision but the explanation doesn't hold together: at that time of the year the wind doesn't blow strongly enough to carry the ships forward before midday and Antony, who had been in the area for months, must have been aware of the fact. No doubt he intended to start waging battle in the morning and, at midday, when the wind got up, to "take off" and head for the open sea; that is why -contrary to what was customary- he had had the sails taken on board so that no enemy vessel could escape him. Thus it was not on purpose or due to either of the two adversaries that the battle only started after midday and that it lasted so long that, according to Suetonius, Octavian was unable to disembark and had to spend the night on board.

Consequently, this raises two questions. The first one, we have just asked: why did Antony wait three hours before opening hostilities? The second one is: why didn't he attempt to ram the enemy? Why didn't he resort to using the weapon of choice in such cases? The huge *rostra* on his powerful vessels would have crushed the hulls of the enemy ships, most of which were of an inferior tonnage. Instead of that, he opted for the use of projectiles and the tactic of trying to board the enemy; his ships drew slowly forward until they were within fighting distance of his adversary's fleet.

Here is what Plutarch writes: « *Though the struggle was beginning to be a close range, the ships did not ram or crush one another at all, since* **Antony's owing to their weight, had no impetus, which chiefly gives effect to the blows of the beaks***, while Caesar's (Octavian) not only avoided dashing front to front against rough and hard bronze armour, but did not even venture to ram the enemy's ships on the side.* (2) » This explanation which we have already mentioned -see above- does not stand up under close scrutiny: in a frontal attack, the greater the bulk of the moving ship, the more serious the damage inflicted on the enemy . . . unless the frontal attack is handicapped not by the weight of the vessels but by the fact that they are unable to acquire enough impetus for the blows to be effective. But, when the enemy line is one and a half kilometers away, there is plenty of scope to reach the necessary speed, approximately 9 knots. Octavian's tactics are -given the lesser tonnage of his ships- as logical as Antony's are incomprehensible. Because, if we are to believe Plutarch and Dio Cassius, Antony deliberately chose to keep his vessels at a standstill and then to use projectiles and try and board the enemy. As historians in antiquity are wont to do, they fictitiously recreate Antony's and Octavian's speeches before the battle. Here is what Plutarch has the former say: « *Antony visited all his ships* **in a row-boat,** *exhorting the soldiers,* **owing to the weight of their ships, to fight without changing their position, as if they were on land. He also ordered the masters of the ships to receive the attacks of the enemy as if their ships were lying quietly at anchor, and to maintain their position at the mouth of the gulf, which was narrow and difficult.** (3) »

Dio Cassius has him make a similar sort of speech: « *See* **the length and beam** *of our vessels, which are such that even if the enemy's were a match for them in number, yet because of these advantages on our side they could do no damage either by charging bows-on or by ramming our sides. For in the one case* **the thickness of our timbers,** *and in the other* **the very height of our ships,** *would certainly check them…* »

As for Octavian, this is what Dio Cassius has him say: « *Will they not by* **their very height and staunchness be more difficult for their rowers to move and less obedient for their pilots?** (4) »

But -let us say it once again- these are fictitious speeches and their function is none other than to cover up an incomprehensible anomaly in Antony's choice of tactics: because his ships were much bigger in size, he should have opted for a frontal attack -charging bows- and ramming the enemy- which would have been more likely to succeed and more deadly.

Modern historians have paid little attention to this particular aspect. For the most part they have accepted the two explanations given by ancient sources for Antony's absurd tactical decision:

1. Antony's crews were not experienced enough. This is exactly why Antony shouldn't have chosen to fight at close quarters and to try and board his adversary's ships; it is the sort of tactic that requires quick and complex maneuvers in order to taunt or to avoid the enemy who, actually, turned out to be much more proficient at it.

2. Antony's ships were slow, heavy and unwieldy monsters. This statement is very likely to have been far from true although it has always been universally relayed. Florus, for instance, wrote:

« [Antony's ships] « *having from 6 banks of oars to 9, and being mounted with towers and high decks, they moved along like castles and cities, while* **the sea groaned and the winds were fatigued. Yet their magnitude was their destruction.** (5) ».

Yet the *rostra* which have been excavated by archeologists in various parts of the Mediterranean and a study of the cavities in the *rostra* from Antony's vessels which adorn the *Tropaeum* erected by Octavian at Actium have revealed that the difference between triremes and decaremes is not mathematically proportionate. Thus, a decareme was not three times as long as a trireme but only twice as long, approximately. The difference resided mostly in the tonnage and so in the draught. Therefore, contrary to legend, Antony's largest vessels were not great monsters which were impossible to maneuver, even if they were less easy to move and turn than triremes or *liburnae*.

This legend dates back to Antiquity. How did it originate? Its roots are certainly to be found in Augustan propaganda whose aim was to stress that victory over vessels presented as monstrous sea creatures had been obtained by ships built by and for men. But the primary reason why it was so widely believed is that in Plutarch's, Florus's and Dio Cassius's day -one or two centuries after the battle- big ships like those with 6 or 10 rowers per bench- hadn't been built in a long time. In fact, they hadn't been built since Actium because the Mediterranean -*mare nostrum*-, which was now at peace only required the attention of a "maritime police" made up of much smaller vessels. For Pliny, Vitruvius and Vegetius, living under the Emperors, the quinqueremes are the biggest ships there are. Doubtless, Antony's largest vessels were, despite their size, quite effective in battle.

Out of all this, two bare facts are worth noting: 1. *Antony's ships remained motionless for a long time*. 2. *Then, they were difficult to move*. Why? Two details in a text which I have already quoted -see above- cannot fail to intrigue: 1. Why does Antony visit his ships in a row-boat in order to exhort his troops instead of cruising in front of his fleet in his command decareme? Indeed, it was contrary to common practice in ancient times for a leader to address his soldiers from below. It is as if Antony had been unable to use his flagship to move about. 2. Why does Plutarch have him tell his sailors to "*mind the difficult mouth of the gulf* "(6)? What particular dangers could have lurked in this narrow channel that ships have to go through to leave the Ambracian Gulf and reach the open sea?

The elements presented *supra* do not enable us to answer these last two questions nor those previously asked about the reason for Antony's ships being at a standstill and for his absurd tactics. Is it possible to go any further?

It is necessary here to add to the file two pieces of evidence, which have so far gone more or less unnoticed. The first is a passage by the poet Propertius, a contemporary at the time of Actium. In an elegy written in 16 BC, he recalls this battle, which became the basis on which Augustus built his regime. Before the battle, Apollo is supposed to be addressing Augustus (Octavian) thus: « *Do not fear that their ships are winged with a hundred oars:* **their fleet rides an unwilling sea.** (7) »

The language is certainly poetic with the oars being compared to wings but the main point, that which we must remember, is that the sea *is unwilling* to let Antony's fleet ride it.

The other document to be added to the file is much more telling and it is quite surprising that it isn't mentioned more often. It is a passage by Dio Cassius: « *When they (Antony's soldiers) set sail at the sound of the trumpet, and with their ships in dense array drew up their line* **a little outside the strait and advanced no further**, *Caesar (Octavian) set out as if to engage with them, if they stood their ground, or even to make them retire. But when they* **neither came out against him on their side nor turned to retire, but remained where they were, and not only that, but also vastly increased the density of their line by their close formation, Caesar checked his course, in doubt what to do.** *He then ordered his sailors to let their oars rest in the water, and waited for a time; after this he suddenly, at a given signal, led forward both the wings and bent his line in the form of a crescent, hoping to surround the enemy, or otherwise to breach their formation in any case. Antony, accordingly, fearing this flanking and encircling movement, advanced to meet it* **as best he could,** *and thus reluctantly joined battle with Caesar* (8) »

This account is much more detailed than Plutarch's description of the same stage of the battle. It is also slightly different, probably because the original source is not the same. What we learn from it is that hardly had Antony's vessels come out of the narrows that they stopped moving, causing a "bottleneck" behind them. This unexpected turn of events surprised Octavian. Plutarch corroborates this. Then, according to Dio Cassius, it was Octavian who was responsible for engaging; according to Plutarch, those responsible for engaging were Antony's troops because they were eager to fight. But, in any case, it was never Antony, Antony who was unwillingly compelled to do battle in conditions described as being less than favorable. In this passage, there is no mention of a deliberate tactical choice on Antony's part to try and board the enemy; the way the battle was waged was obviously dictated by outside circumstances with Antony behaving as if he were paralyzed.

To conclude, it seems very clear, after examining ancient sources, that *something happened* which prevented Antony from launching, as was expected, his fleet against the adversary's fleet, taking advantage of the greater bulk of his ships to ram the enemy. Instead of which, he was forced, first of all, to remain *for a long time at a standstill*, to the great surprise of his adversaries, letting his vessels form a bottleneck behind his frontal line. And then, afterwards, compelled by the enemy to do battle, he advanced slowly towards them, which made ramming impossible. All that remained for him to do, then, was to get close to his opponent's vessels and try and board them, hindered though he was by the sheer size of his ships. Contrary to what our sources would have us believe, he did not *choose* these tactics, but was compelled to use them for some unknown reason. The aim of this study is, indeed, to try to find out the truth about this unknown reason.

1. Plutarch, *Life of Antony, 65*, 6-8*;* cf. Cassius Dio, *Roman History,* 50, 23, 2
2. Plutarch, *Life of* Antony*,* 66, 1
3. Plutarch, *Life of* Antony*,* 65, 4
4. Cassius Dio, *Roman History,* 50, 29, 2
5. Florus, *History of the Roman people, from Romulus to Augustus,* 2, 21, 5
6. Plutarch, *Life of* Antony*,* 65, 4
7. Propertius, *Elegies,* 2, 16, 37-38; cf. 4, 6, 47-48
8. Cassius Dio, *Roman History,* 50, 31, 3-5

## 9.2   SUPPLEMENTARY INFORMATION ON LINGUISTICS

For several centuries, from Ancient Greece until the 16th century, it was said, and repeatedly so, that a small fish called echeneis had the magic power of holding back ships when it latched onto their hulls. Nineteenth century naturalists, such as B. G. de Lacépède, took a rather caustic view of this ancient tale: "From the days of Aristotle until the present this animal has been the object of constant attention; its shape has been examined, its habits have been observed and its physical characteristics have been scrutinized. Not only was it considered to possess magic properties, absurd abilities and ridiculous strength, but it was viewed as a striking example of the occult qualities dispensed by nature to its offspring. It appeared as a convincing proof of nature's qualities, secret in their origins and in essence unknowable; the fish was honored in the imagery of poets, in the analogies of orators, in travelers' narratives and in naturalists' descriptions. [...] How many fables and errors have been accumulated in such passages, which are also stylistic masterpieces?" (1) Nowadays, the spontaneous reaction is indeed to wonder how people could have believed in such improbable powers for so many centuries. One of the explanations is perhaps to be found in the argument put forward by the historian M. Bloch, (2) who takes the example of the remora (the Latin name of the echeneis) to demonstrate how, before the development of critical methods of checking information and witnesses in the reconstruction of history, the most intelligent minds accepted a given fact without questioning its veracity. It was based on tradition, all the more so when those "facts" were handed down by renowned minds of the past. But while this may explain the transmission of the legend, it does not solve the mystery of its origin: where did the legend begin and when? What exactly was said about the echeneis in Ancient times? When was the fish first considered to possess such extraordinary powers? **It is our intention to explore the context in which the legend was born through rigorous analysis of Ancient sources**, in order to better understand the link between beliefs and their unavoidable imaginary elements, and knowledge: the facts and discourse which they gave rise to, the level of knowledge and, perhaps, their exploitation by the elite.

Research into the occurrences of the substantive noun *echeneis* in Greek literature, and its Latin translations in the terms *mora,* and *remora*, also including the variants *remeligo,* and *remirora*, reveals the relative rarity of such texts: **only fifteen or so authors in a corpus which covers a period from the 5[th] century B.C. to the 7[th] century A.D**. The fish is mentioned in various works: naturalists' descriptions of fish in natural history treatises or in didactic works (Aristotle, Ovid, Aelian and Oppian (3)); poetic embellishment (in Lucan's writings, (4) for instance, where it appears as an ingredient in a magic potion); in a chapter from an encyclopaedia (Pliny the Elder (5) combines a descriptive passage of the fish with historical anecdotes and a commentary on its associated magical properties); a banquet anecdote (Plutarch 6); an allusion in the letters of a Roman statesman (Cassiodorus 7); in the writings of later commentators (grammarians or Church Fathers eager to explain the complexity of the world through etymological explanations, or through the collection of pagan knowledge which was reinterpreted in the light of the greatness of the divine: Donatius, Servius, Isadore of Seville and Ambrose 8). In addition to the noun *echeneis*, the adjective *echeneis, idos* which carries the meaning "which stops or holds back vessels", (in Aeschylus during the 5[th] century B.C., Nonnos of Panopolis during the 5[th] century A.D. and Theaetetus Scholasticus during the 6[th] century A.D. 9).

This quick survey shows that the occurrences are somewhat marginal. **The most defining text is without doubt that written by Pliny the Elder a few years after the Battle of Actium,** in which he insists on the strange immobilization of Mark Antony's fleet during that historical event, attributing the cause to the powerful action of the fish to which he gives the name *mora*, a noun which also means "a delay, or lateness" in Latin. Yet there are two striking elements: firstly, among Ancient historians who provide a detailed description of the Battle of Actium (Plutarch, Orosius Florus, Dion Cassius), none explicitly mention the fish, not even Plutarch who refers to it in his *Table Talk* but not in his *Life of Mark Antony*; secondly, there is a concentration of occurrences in the 1st century A.D., as regards the longer descriptions (Ovid, Lucan, Pliny, Plutarch 10). **The "legend of the echeneis" would appear to be a relatively late construction**, containing several strata, the earliest of which goes back to Aristotle, with more frequent references clearly appearing from the 1st century A.D. onwards, in other words around the same time as the Battle of Actium. Aristotle does not indicate that the fish is able to hold back ships; historians writing before or at the same time as the battle between Mark Antony and Octavian do not see fit to mention it either. It is therefore tempting to postulate that the legend was born during the Roman recovery of a Greek belief (mainly transmitted through oral tradition) in the magic powers of the fish (Aristotle describes its use in making magic potions and to delay court trials and slow down justice), to which the Romans gave a new lease of life based on the events at Actium, in order to increase the marvellous powers of the fish.

One way of disentangling the skein of suppositions is to carry out an analysis of the discourses which accompany references to the fish and its exceptional powers throughout classical and late Antiquity. These discourses are clustered around five poles which need to be considered in greater detail:
magic, nature, reason, religion and politics.

9.2 (a) Magic power.

Chronologically, the first text which has been preserved is Aristotle's (11) which mentions the use of the fish to slow down court trials and in the making of potions. Lucan (12) mentions it as an ingredient for a resurrection spell in the description of witchcraft in Book 6 of the *Pharsalia*. Pliny (13) refers to the belief held by some Greeks that if worn as an amulet the fish prevents miscarriages or favours delivery at childbirth (in which case it is given the name *odinolytes*). Pliny places the fish among the list of antiaphrodisiacs, (i.e. reducing amorous passion), along with "rhinoceros skin taken from the left forehead and attached in a lamb's skin". He also repeats Aristotle's indications of its use for the making of potions (and more specifically erotic potions) and for court trials (to slow them down). We may therefore suppose that there was an idea of "mimetic" functioning despite the fact that the texts make no specific reference to this idea, and there is no indication of whether there was a *magus* involved in ordering the action through magic formulas which might use the fish symbolically to obtain this effect, or whether the power of the fish is attributed to its physical characteristics, say of a magnetic nature. Might it be the case that people believed, by analogy, that the fish also had the extraordinary power to hold back ships to the extent that it could even slow them down to a complete standstill? The only document in which the holding back of a ship is explicitly linked to magic intervention is in one of the later texts, the *Cyranides*, a compilation of works on magic written between the 1st and the 8th centuries A.D., in which it is claimed that if just a few echeneis bones are sewn into horse leather and then brought aboard a ship, hidden in clothes, then the ship will not be able to move forward (14).

We should not judge too quickly. Beyond the folklore of oral beliefs or popular traditions, magic did have a very real impact on people's attitudes. The long defensive speech from the 2nd century A.D. in which Apuleius denies the charge of having charmed his wife by the use of magic, notably through the administering of a fish-based potion, provides sufficient proof of such beliefs (15). Even Pliny, who shows himself to be sceptical as regards magic and who is keen to demystify the sham of magic at a time of firm belief, nevertheless describes some strange recipes, such as attaching a bramble-frog to the body in a piece of fresh sheep-skin, in order to put an end to love (16). Closer to our own subject, he also indicates that a boat can be held back by a no less irrational expedient: bringing the right foot of a tortoise on board (17) !

We need to return to one decisive element: the fish attaching itself to the boat. The remora certainly does attach itself to surfaces using a flat disc on its head which has cartilage blades which act as a sucker. By creating a vacuum between these blades, or by hooking the spines which cover the rear edge of the blades, the fish attaches itself to rocks, boats or to other creatures. Quite understandably, all the authors insist on the fish's ability to attach itself. Yet the curious thing is that the sole ability to attach itself seems to be used to explain the slowing down, or even the stopping of a ship. Pliny implicitly suggests there is an immediate effect on a storm when the fish attaches itself to a vessel ("It easily puts an end to force and tames the fury of the elements, effortlessly, merely by attaching itself" (18)). Is this merely a form of poetic hyperbole, which takes pleasure in developing an *adunaton*? Isidore repeats Pliny, almost word for word: "The ship seems to behave as if it were anchored to the sea and remains motionless, not because the fish is holding it back but because it has attached itself to the ship" (19). This implies that the fish does not hold the ship back as such, but rather its astonishing action is revealed only by its attaching itself. The key term which is repeated in almost all of the texts is *adhaerere*. An accurate interpretation is essential here because this is where we may understand what the authors intend by the term "adhere": of course this means first and foremost that the fish is attached and cannot be detached; yet this single element appears to explain the cause and effect relationship between the fish attaching itself to the surface and the resulting immobilization of the object, without raising any further questions.

According to L. C. Watson (20), *adhaerere* corresponds to the Greek verb *kollô* "to stick", common in Greek magical papyrus which describe love charm rituals. This "sticking" creates a physical "link", the equivalent of *katadesmos* of love charms. Indeed, the terms used by several authors (regarding the constraint, the obedience, the preventing and the link with the boat) suggest a magical connection (cf. Oppian (Hal. 1, 232-3 *ouk etelousa, pepedètai; 235-6 desma; 242-3 pedèn*); Aelian (N.A., 2, 17 *pedèsas*); Nonnos (Dion. 21, 45-8 *katasketon desmo* et 36, 367-9

*desmo*; Pliny H.N. 32, 2-6, *tenere uincta*). If our analysis is accurate then this signifies that it was believed that a boat could be immobilised in the same way that a man could be linked to a woman in an erotic context. Cassiodorus, who does not believe in the fish's power whatsoever, uses the verb *adligare* (21) to describe the way in which the echeneis bites the sides of the ship; yet this is the verb which is used four times by Pliny in the form of the participle *adalligatus,* (22) to designate the wearing of an object in the shape of an amulet in a magical context. He also says that the boat seems to be stuck (*infixum*(23)) to the surface of the sea.

**These examples show that a specific vocabulary with a high degree of magical connotation was projected onto the fish**, probably derived from the Greek beliefs in magic which are attested to by Aristotle — but this does not necessarily imply that the authors who use such terms actually believe in any magical power. This point therefore needs to be examined more closely. The notion of a magic link does at any rate explain what Pliny presents as an incongruous detail: when an echeneis was found under Caligula's ship, there was general surprise at the fact that the fish no longer had any effect when it was taken on board, as if any slowing down effect due to mechanical force or traction was out of the question. (24)

9.2 (b)   One of the wonders of nature.
When Pliny mentions the astonishing characteristics of a fish able to hold back a vessel, he sees this above all as an irrefutable indication of the mysterious power of nature. The action of a small fish which is able to resist the fury of the elements leads to a consideration of the theme of nature triumphing against itself after a struggle between antagonistic forces. As we have seen above, he provides no explanation for this power, merely presenting it as a fact of nature, proven by observation which is sufficient to validate that fact. Two historical anecdotes are used to support this assertion: first of all, the Battle of Actium, and secondly the immobilization of Caligula's ship during a voyage he undertook between two Latium coastal towns. Pliny provides no further analysis on how the fish functions and concludes his description with a general formula, widening his demonstration to include a broader group than the echeneis species: "there is no doubt in our minds that these animals [in other words all the astonishing creatures produced by nature] have identical powers". To illustrate this power, he quotes a remarkable precedent to be found in the similar action of sea shells which stopped a Greek expedition during the time of Periande: these were marine gastropod molluscs called Venus shells which have a porcelain-like inner layer.

**This belief in the power of nature follows a line from the paradoxography works** which flourished in Greece from the Hellenistic period onwards. The authors of these works applied themselves to compiling natural wonders, attempting to astonish or amuse their readers, through exotic or sensational descriptions. Viewed from this perspective the prowess of the echeneis is no more extraordinary than that of the phoenix, the unicorn or the basilisk, and it is not unusual to read surprising stories such as that of the literate pachyderms who can read Greek (25) or that of the pilfering octopuses which climb trees to steal fruit (26). Such anecdotes correspond to the taste of their readers who were keen on such curiosities, as is revealed by the development of the notion of prodigious feats to be described below.

9.2 (c)   Rational interpretations.
It might be expected that the appeal to common sense might prevail, or at least be well represented, but this is not the case. Rational interpretations are in the minority. In *Table Talk*, (27) Plutarch explains that boats slow down because of the algae which build up on the hull and the rudder, especially when the boat has been at anchor for a long period of time. The keel then becomes gorged with water and therefore accumulates a large amount of algae, the wood becoming covered with moss and losing its power of penetration in the water, while the waves which strike this sticky mass do not bounce off it effectively. Plutarch, who was aware of the phenomenon of magnetics, clearly excludes the latter explanation, which is put forward by one of his guests in an attempt to call on common sense and deconstruct the legend by reversing the relationship of cause and effect: he suggests that it is the presence of algae which attracts the fish and not the fish attached to the boat which slow it down. The idea of a whole shoal of fish having an influence on the advance of a ship might seem slightly plausible but this hypothesis is never suggested by the Ancient texts, contrary to the Renaissance emblems in which clouds of sucker fish appear attached to the keels of boats. Five centuries after Plutarch, one of Cassiodorus's letters suggests human causes: the late arrival of boats loaded with important wheat cargoes was not due to the fantastical effect of an echeneis but rather to the negligence of sailors who may have fallen asleep, or who simply did not care. The humour and cultured elegance of the statesman is combined with moral judgement: "the echeneis which slows them down is their own venality, the conch stings, it is their own unlimited passions". (28)

9.2 (d)   A divine warning.
Pliny claims the echeneis is: "a fish which is worthy of counting among the omens", (29) thus superimposing the religious prism over a phenomenon which thus acquires a new dimension. Roman religion cannot be invoked without mentioning the importance of monstra, the spectacular signs which it was believed were sent by the Gods in order to warn mankind that there was a message for them. The monstrum designated both the phenomenon of the apparition which revealed divine will and, at the same time, the exceptional creatures which transmitted this sign. In our case this is a strange fish endowed with supernatural powers. The echeneis does indeed belong to the family of "monsters", and, to be more precise, it is a prodigious animal: it is thus to be distinguished from the monstrous teratological creature such as the four-legged snake; it is also to be distinguished from what we could term "prodigial" animals, that is to say ordinary animals which suddenly behave in a strange way thus announcing a divine message, such as a bird stealing embers from a bonfire. The prodigious characteristic of the echeneis is its species' singular and intrinsic ability to hold back boats, as is repeated in Ancient etymologies. Another surprising aspect of this fish may be added: in the impressive list of prodigious animals drawn up by Julius Obsequens only two involve fish. The act of one single fish such as the echeneis is thus exceptional.

Pliny considers the echeneis to be a bad omen because it announced Caligula's death. Pliny's testimony reveals the transformation of the echeneis as a Roman prodigy during the Roman Empire: firstly, the prodigy, which was initially a sign of divine intervention in human affairs, had evolved to become much more an omen, of a divine nature. Secondly, the prodigious fish became more specific, no longer necessarily referring to a group but rather to the specific destiny of one individual. The anecdote of the role of the echeneis at Actium thus takes its place in a series of omens which announce the defeat of Mark Antony, revealing to the eyes of the world that nature and the Gods had chosen that day to side with the Octavians. A linguistic detail may corroborate this reading: the Latin term mora, and its synonym remora, which came to replace the former, present assonance with an Ancient term connected with omens, remur, which designated a bird of ill omen. It is therefore possible that the remora may have sounded as if it carried negative connotations.

In late Antiquity, Oppian (30) also describes the fish as prodigious, but in a figurative sense, implying that it was among the prodigious visions produced by dreams, with a corresponding loss of its divine quality. Yet Isidore of Seville and Ambrose reintroduce the religious interpretation by detecting a reflection of the Creator's omnipotence in the extraordinary power of the creature. "Do you think that so much power has been given to it without a gift from the Creator?", (31) writes Ambrose, who uses the example to demonstrate that a fish such as the echeneis is used to remind mankind of its limits and of its condition, by placing it in a situation in which it can only expect help and safety from the Lord when faced with the perils of life.

9.2 (e)   Political exploitation.

As has just been suggested, the appearance of signs and their interpretation is always of interest for political leaders. One of our hypotheses is the following: the victors of war, Octavian and his followers, may have spread the rumour of the intervention of the echeneis for propaganda purposes, in order to prove that the Gods had decided to side with them. Such political exploitation of religious beliefs was not new. The great political events of the end of the Republic were accompanied by prodigious events which poets and historians had busily chronicled: rains of blood, rivers reversing their flow, statues of the Gods covered in sweat, a mother giving birth to a snake... During the transition between the Republic and the Empire, towards the end of the 1st century B.C., the historiographical tradition reports a number of omens regarding Octavian-Augustus and the imperial family, destined to an exceptional fate: Octavian was born of the union between Atia and Apollo-snake; (32) an eagle is said to have stolen a piece of bread from him and then returned it to him, a sign of his future sovereignty; (33) Livia is said to have warmed an egg in her hands hatching a chick with a huge crest, thus announcing prosperous offspring and the gaining of power as represented by the crown symbolism of the crest.

During the period immediately preceding the Battle of Actium, a long series of prodigies were reported, often involving animals. To take the example of one single historian, Dion Cassius describes how, in 36 B.C., a fish jumped from the sea to the feet of Sextus Pompeius, and the diviners told him that he would be master of the seas; (34) in 32 B.C., (35) a monkey interrupted a ceremony in the Temple of Ceres, a victory statue fell on the stage of a theatre, Etna erupted, a two-headed snake appeared in Etruria eventually to be struck by lightning, a statue of Mark Antony wept floods of tears, a wolf entered the Temple of Fortune, a dog was devoured by another dog during a horse race in the circus; in 31 B.C., just before Actium, Cleopatra fretted about swallows nesting on her admiral's ship, lightning knocked down the statues of Mark Antony and Cleopatra erected by the Athenians; (36) and then it rained blood, weapons appeared in the sky, drums and flutes were heard, a giant snake appeared, the statue of Apis began lowing and comets were seen….(37) In this context the attributing of the incomprehensible immobilisation of Mark Antony's vessel to the action of an echeneis and the view that this implied that the Gods had intervened in human affairs thus constituted a perfectly plausible hypothesis. The legend may have begun just after Actium, during the ten-ship dedication ceremony which was offered by the victors just after the battle. Only the Octavians could have participated in hauling the ship out of the water. The legend may have originated "from the fact that when the ship of the defeated admiral was hauled on land to be exhibited as a trophy, a remora type fish was discovered on the hull". (38) Mark Antony's boat had spent time in the waters of the Bay of Preveza and Vonitza and the hull was probably laden down by parasitic plants and animals, and it is plausible to suggest that this may have attracted fish. The witnesses to the ceremony probably associated the immobilisation of the boat to the presence of one or more fish parasites, due to the widespread belief in the Mediterranean of the immobilising power of certain sea shells or fish. (39). These beliefs may have been seized upon by the Octavians to crown their victory with divine support. (40)

This anecdote is also present in Octavian's Neptune like propaganda after Actium: he claimed to have been given mastery over the seas and that this was proof of his divine election. Used for ideological purposes, the legend of the remora suggested through its imagery that the powerful Mark Antony, his vessels like monstrous centaurs, could do nothing against the will of the Gods, who could brandish a tiny fish to put a permanent end to his advance. Conversely, a description by Propertius represents Octavian as the worthy protégé of Apollo, (41) who appeared over the stern of his boat surrounded by a triple flame. A gloss by Servius (43) commenting on one of Virgil's lines suggests that the exegetes of late Antiquity thought that the fish had been sent by Neptune to hold back Mark Antony.

After a long period during which the echeneis was purely considered from ichthyological perspectives, critical discourse on the fish is today mostly the fruit of research by researchers in the arts and humanities. The story of the prodigious little fish and the commentaries which it gave rise to has led to four main approaches: linguistics, mythology (more specifically the mystification of history), the transmission of texts and the influence of the legend in European literature.

As regards the linguistic approach, the Ancient authors named the fish after the legend it is associated with, considering from the start that the name derived from the creature's powers. Contemporary linguists partly confirm this interpretation. They analyse the noun as a zoonym made up of two juxtaposed radicals, a verb (*echein* to hold, to hold on to) and a noun (*naus* the ship), associated in a noun which does not reveal the syntactic relationship which unites the two radicals, as is usual in this type of compound. If we look for the implicit sentence which would provide the semantic base for the creation of this noun there are two possibilities due to the fact that the verb may function either transitively or intransitively: "he holds on to the boat" or "he holds back the boat". This is where the ancient etymology (the fish was given this name because it holds back the boat) differs from contemporary linguistic reinterpretation (the compound may also signify that fish holds on to the boat). **The modern day supposition is that the animal "which attaches itself to hulls" (an intransitive construction which probably initially led to the term in Greece) was later perceived, at a time and period still to be determined, as the animal "which holds back boats" (transitive construction)**. The legend may therefore originate in an etymological shift.

A second direction concerns the spread of the legend and its relations with the field of myth. As Pastoureau has written, the collective imagination of a period allows us to understand that period as surely as the events which took place and the prevalent living conditions: "The historian must never excessively oppose imagination and reality. For the historian, as for the ethnologist, the anthropologist or the sociologist, imagination is always part of reality." (43) To take into account this imagination involves close study of a specific cultural context and reasoning within the Ancients' representation of the world. A reconstitution of the legend therefore involves the job of discriminating between what has been observed, believed, thought and imagined. Knowledge 2,000 years ago was considerably different from ours today, even in the field of an apparently accessible field such as zoology: **people were able to believe in a fish with supernatural powers in the same way that they believed in the existence of fantastic creatures such as the griffin, the phoenix, the unicorn, the manticore and the amphisbaena**, or more extraordinary still, in the metamorphosis of storks into women in the Oceanid islands. (44) The frontier between fable and reality was thus a moveable feast. Aristotle confirms the existence of a lemnian billy goat with two udders near its penis which were milked to make cheese. The same is true of **monsters, the cynocephalus, hermaphrodite foal and the hippocentaur**. The echeneis also needs to be considered alongside **imaginary marine creatures, mermaids, tritons, Nereids, Charybdis, Scylla**, whales, swordfish and all the dangers that are supposed to inhabit the troubled depths of the subaquatic world. It must lastly be viewed in relation to a whole **bestiary connected with the exercise of power** (Augustus' parrot, the salt fish caught by Mark Antony, Cleopatra's viper and dissolved pearls), their fantastic nature being heightened by the fantastic habits believed to be widespread in a mythic and sulphurous East.

If the legend was handed down through the centuries from Classical Antiquity until the Renaissance, it is without doubt due to a process which needs to be fully explored: that process is based on a **respect for tradition, which upholds the supremacy of text**. Trust in the authority of a source sometimes annihilated all critical thinking. Over a long period "any fact which was claimed in writing, was, three-quarters of the time, accepted as fact." (45) This form of unquestioning transmission is demonstrated by the filiation which may be observed between Aristotle, Pliny who translates the former, Cassiodorus and Isidore of Seville who quote the Latin encyclopaedist almost word for word, and the French texts of the 16[th] century which translated them in turn. From this perspective, the texts are not only to be viewed as proof of what was said and thought, but also as having generated discourse and reactions to such discourse. When the text becomes a reality in itself, the story of a text sometimes ends up replacing reality.

Lastly, the circumstances surrounding the **spread of the remora legend in European literature** need to be retraced, as well as the place of the legend in medieval bestiaries, alchemy, the marked upturn in interest for fish in the 16[th] century, probably due to two historic events: the Battle of Preveza and the crossing of the Cardinal of Tournon (Francisco Massari, Edward Wotton, Rabelais, Conrad Gesner, Jérôme Cardan, Rondelet, Alciat, Ambroise Paré and Montaigne), the wind of questioning in the 17[th] century (Kircher, Aldrovandi, Gaspar Schott, Mersènne le Père, François Bernier and Du Tertre), the refuting of its power during the Enlightenment (Diderot, J. Valmont Bomar, C. Favart d'Herbigny and Linnaeus), its metaphoric use during the 19[th] century (Michelet and Balzac), and its legacy in the 20[th] century (Rezvani and Borgès).

## 9.3 SUPPLEMENTARY INFORMATION ON NAVAL ARCHETECTURAL

From historical reports (Pliny the Elder, 1857; Martin, 1995), Octave presented a fleet composed of oared galley (classes from 2 to 6, namely bireme to sexteres). The class refers roughly speaking to the number of rowers per bench (see SI on Mathematics). Unlike Octavian fleet was light, Antonian fleet was heavy with classes from 4 to 10 (class 10 is so-called decareme). As discussed in the supplementary information on Ancient History, the flag-ship of Antony, a decareme, was delayed for several hours whereas the Octavian ships moved freely. Moreover, the contemporary reconstruction of an ancient fifth century BC Athenian trieres by John Coates, John Morrison and Boris Rankov (Morrison, 1996; Morrison et al., 2000; Rankov, 2012) during the 80's and its tests thanks to sea trials since then allowed us to have access to the naval architecture plans of an ancient galley, with the help of the Trireme Trust. In addition, we benefited from the work of William Murray and the Institute for the Visualization of History who provided us with the 3D digitization of a ram, an ancient weapon that equipped the bow of ancient galleys (Murray, 2012). We used the so-called Athlit ram for our reduced model.

Concomitantly to the end of the Hellenistic era, the construction of such big boats was stopped, what Murray calls "the big ship phenomenon". The most delicate point of our study is the following assumption: we chose to consider that the dimensions of a decareme were twice those of a trireme, a strong hypothesis that we will try to justify. Firstly, the ancient reports insist on the gigantism of the biggest boats at Actium: Florus speaks of the Antonian ships "being mounted with towers and high decks, they moved along like castles and cities, while the sea groaned and the winds were fatigued. Yet their magnitude was their destruction". Historians would certainly argue about this point but in absence of direct evidence, we can only make hypotheses (Pitassi, 2011 for a sizes comparison between a 2, 3, 4 and 5 classes). Secondly, thanks to the archaeological studies of William Murray we have indirect evidence of the massiveness of the warships at Actium. Indeed, his team was able to identify the size of the biggest boats thanks to the study of the prints of rams in the sockets of the Apollo temple in Nicopolis. Just after the battle, Octavian dedicated to Apollo a trophy with all the sizes of rams taken on the Antonian boats, from class 1 to class 10. By multiplying the dimensions of a trireme by a factor of two in order to get a decareme, the obtained draft and beam seem to be compatible with the extrapolations of the historians. Legitimate doubts could be formulated with respect to the length since a boat of seventy meters long would maybe imply technological constraints with respect to its building, stability and resistance to flexion. Hence, by doing so, we compensate somehow with the fact that we kept the same block coefficient (the ratio of the box volume occupied by the ship, here 0.37) for the trireme and the decareme. However, as we will see, the important parameter in the context of the naval battle of Actium is not the length of the boat but the respective ships draft (1m for a trireme and 2m for a decareme, see below) versus the water depth: it is very probable that we underestimated the draft of the decareme since 2 m could be increased easily up to 3 m because of the weight of the boats as constantly described by the ancient sources. Finally, we noticed that gigantic boats built for the naumachiae of Caligula in the first century were as long as 74 meters, the so-called Nemi boat (Carlson, 2002) despite the fact that they sailed on a calm lake and not in the Mediterranean Sea.

With interpolation, from historical data on the number of rowers per class galley (Pitassi, 2011, 2012; D'Amato, 2015), we can assume the number of rowers of a decareme (Figure SI1). We find a ratio of 605/170=3.55 rowers between a decareme and a trireme, this is consistent with the resistance ratio.

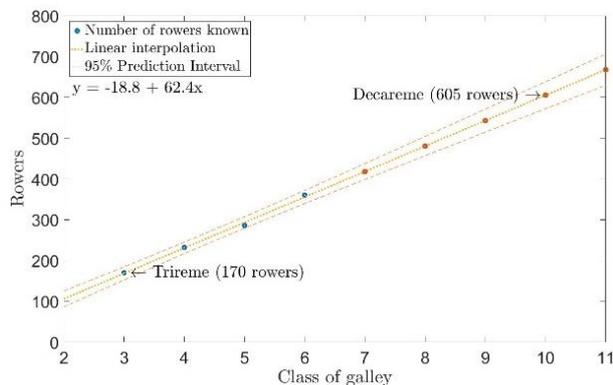

**Figure SI1. Linear interpolation of number of rowers per class of galley.**

9.4 SUPPLEMENTARY INFORMATION ON OCEANOGRAPHY

The Ambracian Gulf is a semi enclosed coastal system in Western Greece, with a mean depth of 26 m and a maximum depth of 63 m. The gulf is connected with the adjacent open Ionian Sea through an elongated and narrow channel, i.e. the Preveza - Actium Straits. Channel's length is about 6km while its width ranges from 0.8 to 2 km. Its eastern part, in the gulf's interior, is about 20m deep and 2.5 km wide, while the channel narrows gradually to the west, with its range reaching 0.8 km in the in the middle. At this extended shallow area, at the entrance of the Ambracian Gulf, the mean depth of which does not exceed 5 m, the battle of Actium took place. Nowadays, in the Ambracian Gulf's sill a navigational channel, of about 13.5 m deep, has been constructed (see Figures SI2 and SI3).

In the maps of Figures 1 and SI3, the reconstruction of the Ambracian Gulf's sill bathymetry, during the period of the Actium naval battle is presented. The main differences between the current and the ancient bathymetry of the area are: a) the artificial channel, which was drained in the 1970's, and b) the mean sea level, which was 75 cm, lower than today. The region where the battle took place, i.e. the gulf's entrance, was very shallow, characterized by a mean depth of about 2.5 m. The depths were progressively increased in both directions, toward the Ionian Sea and the gulf's interior.

A fjord-like water circulation, due to its oceanographic conditions and its morphology (Ferentinos et al., 2010; Kountoura and Zacharias, 2014), characterizes the Ambracian Gulf. Two large rivers, i.e Arachthos and Louros discharge large quantities of freshwater into the Ambracian Gulf (Therianos, 1974), resulting to the ecosystem's permanent water column stratification and to the reduced salinity of the surface layer. This water layer is usually well mixed, and its thickness is typically of the order of a few meters. The pycnocline layer's characteristics (intensity and extent) are spatiotemporally varied, under the influence of seasonal meteorological and hydrological changes in the area. Surface and intermediate (pycnocline layer) waters are freely connected with the open sea through the gulf's mouth. Denser water masses are trapped behind the sill, at the greater depths. Like in most fjord type basins, so in the Ambracian Gulf density variations of the open sea water are crucial for the water exchange, both above and below the sill level (Stigebrandt, 2001). At the entrance to the Ambracian Gulf semi-diurnal tide is prevailed with average range of 5cm and a maximum recorded range of 25 cm$^3$, while at the gulf's interior, the limited fetch of about 35 km$^3$ results to a low energy wave regime.

This study is focused on the gulf's sill area, which is of interest because of the interaction between the gulf's surface brackish water mass and the Ionian Sea's salty waters. This interaction, results to the development of a front, due to the presence of a horizontal salinity and density gradient, which extends from the sill's surface to its bottom. The area's water column behaves like a single layer, while its speed and direction are varied under the influence of wind and tidal phase. Hydrodynamic circulation regime, changes at the deeper parts of the region, where brackish water outflows at the surface and saline water inflows near the seabed, attaining speeds of up to 60 and 80 cm/s according to (Ferentinos et al., 2010). Summarizing, the area of the Ambracian Gulf sill is characterized of great oceanographic interest and many peculiarities, due to its morphology, its location and the interaction of currents, tides and wind.

The objective of the present study is to give a scientific explanation about the Antony's defeat in the Actium naval battle. As the truth is possibly connected with the area's oceanography, it is crucial to answer some questions about: 1) The circulation pattern in the study area today, 2) spatial distribution of the pycnocline in the study area today, 3) The importance of morphology, tide and wind on the area's circulation pattern today and 4) The circulation and stratification pattern of the study area during the battle and their influence to the battle's outcome.

The current hydrodynamic conditions in the area of interest will help us to reproduce the prevailed water circulation during the battle. For this purpose, decisive factors will be the data and information that can be retrieved from the battle description in historical texts. Furthermore, a study of the area's: a) water column physicochemical characteristics, b) currents, c) tidal characteristics and d) meteorological parameters is essential.

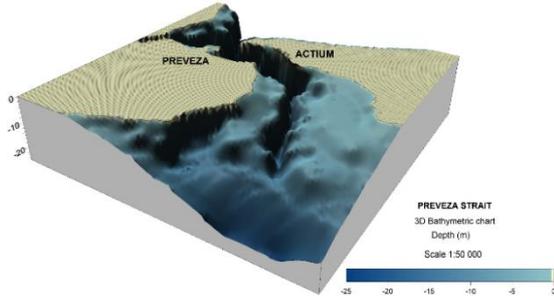

**Figure SI2. Current morphological features of the Ambracian Gulf sill area. 3D block diagram.**

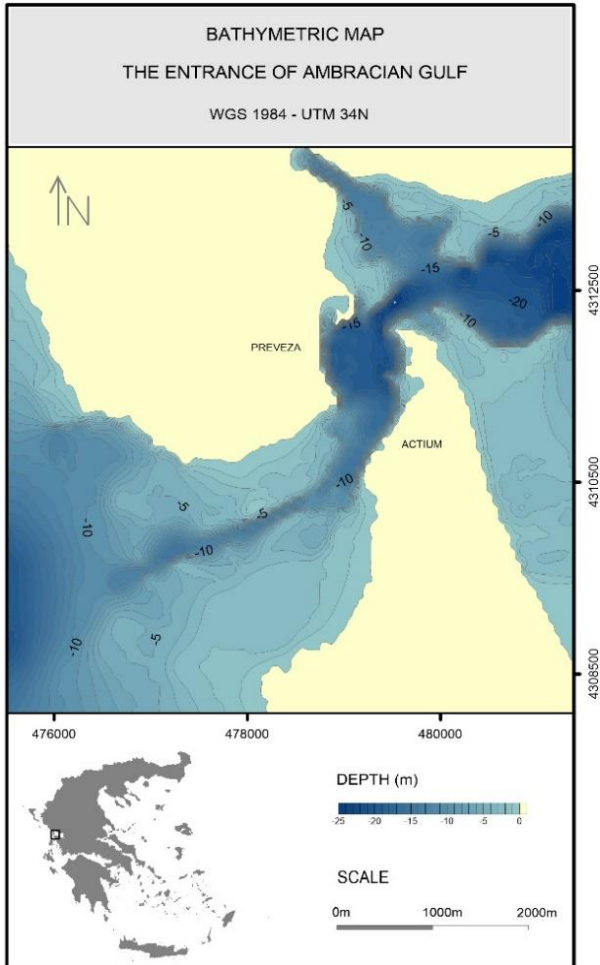

**Figure SI3. The current bathymetric map of the Ambracian Gulf entrance.**

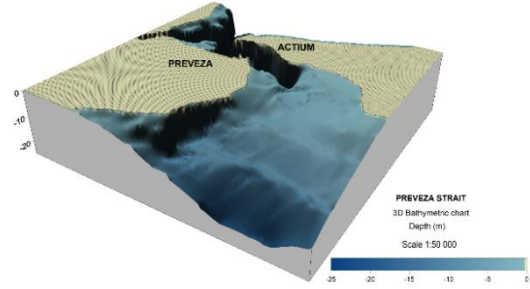

**Figure SI4. Morphological features of the Ambracian Gulf sill area 2000BP. 3D block diagram.**

## 9.5 SUPPLEMENTARY INFORMATION ON MATHEMATICS

The evaluations of the wave-making resistance for the maps is based on an analytical formula called the Sretensky formula (8). This formula is obtained by assuming the ship to be slender, and therefore the wave phenomena to be linear. Many experimental studies (see for instance (2)) have shown a good agreement between the results given by this formula and the data from experiments. This formula takes into account the water depth $h$, the velocity of the ship $V$, and the shape of the ship, given as an offset function f that defines the half-width of the ship for each point of the center-plane $(x,z)$. The direction of motion of the ship is $x$, and the depth is $z$. The wave resistance according to Sretensky's formula hence reads:

$$R_W(U) = \frac{8\pi\rho g}{V^2} \int_{\gamma_0}^{\infty} \frac{I^2(\gamma) + J^2(\gamma)}{\left(\gamma^2 - \frac{g\gamma}{V^2}\tanh(\gamma h)\right)^{\frac{1}{2}}} \gamma d\gamma$$

Where $g = 9.81 \text{m}.s^{-2}$ and the coefficients $I$ and $J$ are given by

$$I(\gamma) = \lambda(\gamma)\frac{U}{2\pi}\int_\Omega \frac{\partial f}{\partial x}(x,z)\frac{\cosh(\gamma(h-z))}{\cosh(\gamma h)}\sin(\lambda(\gamma)x)\,dzdx$$

$$J(\gamma) = \lambda(\gamma)\frac{U}{2\pi}\int_\Omega \frac{\partial f}{\partial x}(x,z)\frac{\cosh(\gamma(h-z))}{\cosh(\gamma h)}\cos(\lambda(\gamma)x)\,dzdx$$

In which $\Omega$ is the domain on which f is defined. The function $\lambda$ is defined by

$$\lambda(\gamma) = \left(\frac{g\gamma}{V^2}\tanh(\gamma h)\right)^{\frac{1}{2}}$$

Finally, $\gamma_0$ is computed as the positive solution of the nonlinear equation

$$\gamma_0 = \frac{g}{V^2}\tanh(\gamma_0 h)$$

whose origin is to be found in the dispersion relation for finite depth water waves (13, 14).

### 9.5 (a) Integration of the naval architectural data

The main advantage of using the aforementioned formula is the opportunity provided to take into account the actual data from naval architecture through the shape of the hull. Recall that the shape of the hull is given by the offset function $f$, in our case, an approximation of it on a mesh. The geometrical data we have on the trireme is not of this form, so our goal will be to recover if though a technique called *draping*. The data we have is in the form of a set of parametric surfaces that can be exploited with the software Rhinoceros.

The principle of the draping technique is to consider a set of points (red points, figure SI5, left) placed initially on a plane parallel to the hull's centerplane, and to project these points on the hull by considering only displacements in the $y$ direction. The set of points we then obtain (red points, figure SI5, right) hence "drape" the object. A few points end up straight on the centerplane, their offset being zero, we trim them off the mesh to avoid unnecessary further calculations.

The draping function is already implemented in Rhinoceros. Three grid of points were used here to represent with the same level of accuracy the hull, the nose and the ram (for which some details have the typical scale of 1cm for a 30m long ship). This results in a mesh file of 413,568 points which is too large for fast and efficient calculations of the wave resistance as it is required here.

We reduce the size of the mesh by extracting a subset of points with a variable density (typically we take more points wherever small details are involved). Then a new mesh is generated with Matlab's build-in Delaunay mesh generator. The result of all these operations is a representation of the trireme's hull through a P1 finite element function on a mesh of 25,185 points (see figure SI6).

This allows us to calculate the functions $I(\gamma)$ and $J(\gamma)$ using a finite element method associated to this P1 representation.

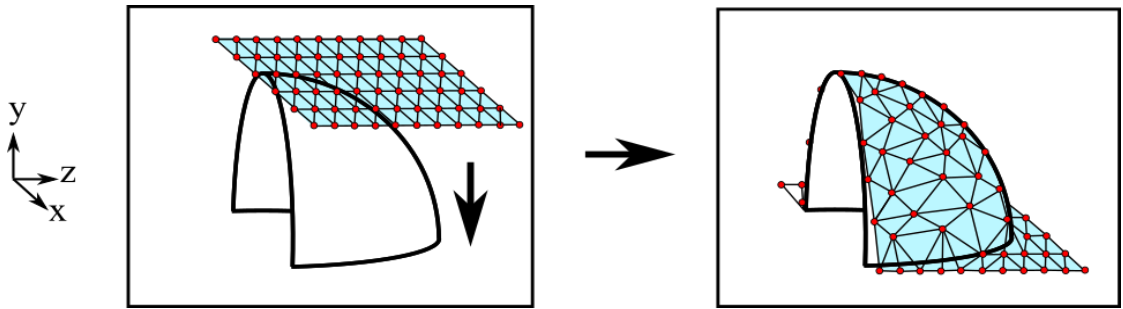

**Figure SI5. Schematic representation of the draping procedure.**

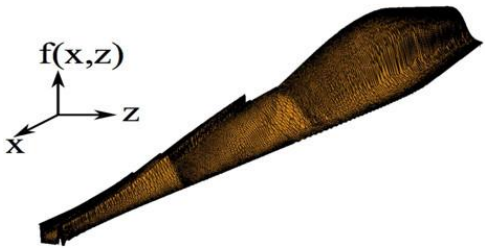

**Figure SI6. Representation of the surface of the function $f$ obtained through our reconstruction procedure.**

9.5 (b)   Calculation of the wave resistance integral

The first difficulty in computing the wave resistance integral (1) remains in the computation of the terms $I$ and $J$, which are integrals with respect to space that depend on the shape of the hull defined by $f$. Let us focus on the computation of $I$ (the computation of $J$ is performed in a similar manner). If we define $\varphi$ as

$$\varphi(\gamma, x, z) = \frac{\partial f}{\partial x}(x, z) \frac{\cosh(\gamma(h-z))}{\cosh(\gamma h)} \sin(\lambda(\gamma) x)$$

our goal is to integrate $\varphi$ with respect to $(x, z)$ for all values of $\gamma$.

The computation of the integrals (1) - (3) is known to be delicate[3]. We detail the numerical method in two steps, first the integrals (2) and (3) with respect the space coordinates $(x, z)$, and then the integral (1) with respect to $\gamma$. The integrals (2) and (3) are computed using the aforementioned P1 representation f of the hull. This term is calculated in an exact manner on every triangle of the mesh by using a mapping on a reference triangle. Such an exact calculation is much preferable to the use of a quadrature formula.

Once $I(\gamma)$ and $J(\gamma)$ are determined, we can compute the integral (1). Two difficulties lie ahead: the singularity at $\gamma_0$ and the infinite range of integration. The first problem is tackled using subintervals that become smaller and smaller as $\gamma$ comes closer to the singularity. The integral at infinity is separated on subintervals that become larger and larger as $\gamma$ grows. On each subinterval, we approach the integral with a two points Gauss integration method. Our resulting numerical wave resistance has been validated by comparison with tabulated results obtained for Wigley hulls.

9.5 (c)   INTEGRATION OF THE OCEANOGRAPHIC DATA

From the methods described above, we are able to compute the wave making resistance for a given $V$, $h$ and $f$. As described in methods section on Oceanography, the bathymetry is given as a 100x74 array that provides a water depth for each point of the Ambracian Gulf. Our aim here is to build an array that provides a wave resistance for each point of the Gulf, for a given velocity $V$ and hull shape $f$. The approach we use in order to reduce the computational costs is to compute wave resistance *vs.* depth profiles and to map these profiles into the oceanographic data by using interpolation. These profiles will be computed with the method described earlier by selecting a limited well-chosen set of depth values ranging from the ship's draught to a depth under which the behavior of the resistance can be considered constant (deep water behavior). The result is then interpolated with splines (Figures SI7 and SI8).

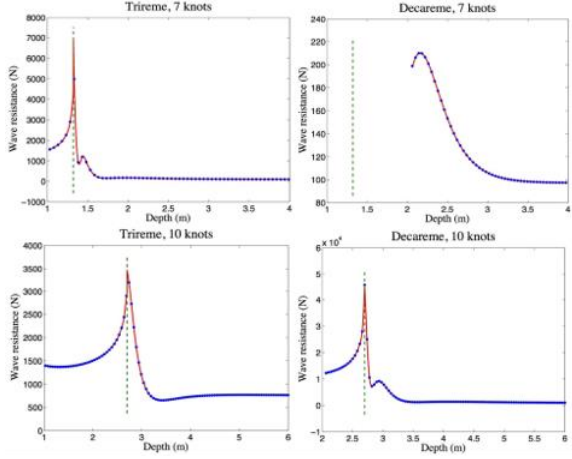

**Figure SI7.** Plots of the wave resistance versus depth profiles. First column: trireme. Second column: decareme. Top row, 7 knots; bottom row: 10 knots. The blue points represent the computed value of the wave making resistance and the red line is the spline interpolation. The vertical green dashed line represents the critical depth corresponding to the critical Froude number $Fr_h = \frac{V}{\sqrt{gh}} = 1$

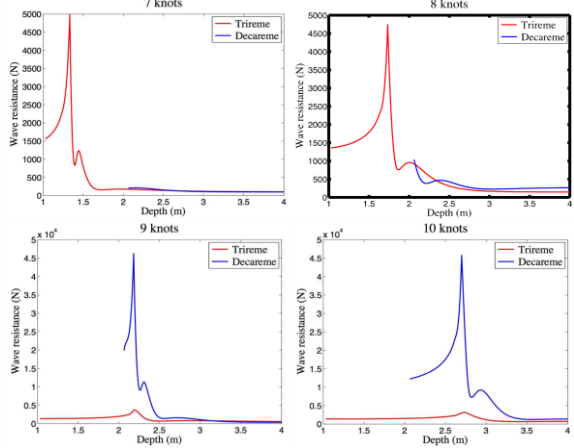

**Figure SI8.** Plots of the wave resistance versus depth profiles for various target velocity from 7 knots to 10 knots. Each plot shows a comparison of the wave-making resistance for each ship involved (in red the trireme, in blue the decareme). Note that the wave-making resistance for depths below the ship's draft (2 m) are not represented as they have no sense. (Gotman, 2002).

water depth for each point of the Ambracian Gulf. Our aim here is to build an array that provides a wave resistance for each point of the Gulf, for a given velocity $V$ and hull shape $f$.

The approach we use in order to reduce the computational costs is to compute wave resistance *vs.* depth profiles and to map these profiles into the oceanographic data by using interpolation. These profiles will be computed with the method described earlier by selecting a limited well-chosen set of depth values ranging from the ship's draught to a depth under which the behavior of the resistance can be considered constant (deep water behavior). The result is then interpolated with splines (Figures SI7 and SI8).

To the wave making resistance, we can add a viscous resistance related to the friction of the fluid on the hull. This viscous contribution can be calculated using the ITTC 57 procedure (ITTC, 1957; Molland et al., 2017):

$$R_v = C_v \frac{1}{2}\rho S V^2$$

With $C_v$ the coefficient of friction:

$$C_F = \frac{0.075}{(\log_{10} Re - 2)^2}$$

Thus, we compute a total resistance, composed of a viscous resistance and a wave making resistance (Figure SI10). For small Froude numbers (equivalent to deep water regime), the wave component becomes negligible compared to the viscous component. In this configuration, the ratio of total resistance is

$R_{tD}/R_{tT} \approx 3.6 \approx Rv_D/Rv_T$ (Figure SI11 and SI12)

This value, close to 4, is explained by the geometry of the decareme which is twice as large as the trireme. Since the galleys have slender shapes, we can relate the ships' geometry to a board of length *L* and height *T*. The wet surface *S* is roughly L × T × 2 (we multiply by two to take into account both sides of the board). The length and the draft of the decareme are double that the length and the draft of the trireme. Thus, there is a factor 4 between the wet surface of decareme and the wet surface of trireme. To find the ratio 3.6, we have to start from the ITTC57 formula:

$R_{vD} = C_{FD}\frac{1}{2}\rho S_D V^2$ and $R_{vT} = C_{FT}\frac{1}{2}\rho S_T V^2$.

$$\frac{R_{vD}}{R_{vT}} = \frac{C_{FD}}{C_{FT}}\frac{S_D}{S_T}$$

$$\frac{C_{FD}}{C_{FT}} = \frac{(\log_{10} Re_T - 2)^2}{(\log_{10} Re_D - 2)^2} = \frac{(\log_{10} Re_T - 2)^2}{(\log_{10} Re_T + \log_{10} 2 - 2)^2}$$

$Re_T \approx 10^8$ so $\frac{C_{vD}}{C_{vT}} \approx 0.9$ and $\frac{R_{vD}}{R_{vT}} \approx \mathbf{3.6}$

Without very shallow effects, a decareme has a viscous resistance 3.6 times stronger than the resistance of a trireme. To compensate this phenomenon, it is necessary to deploy a greater rowing power.
.

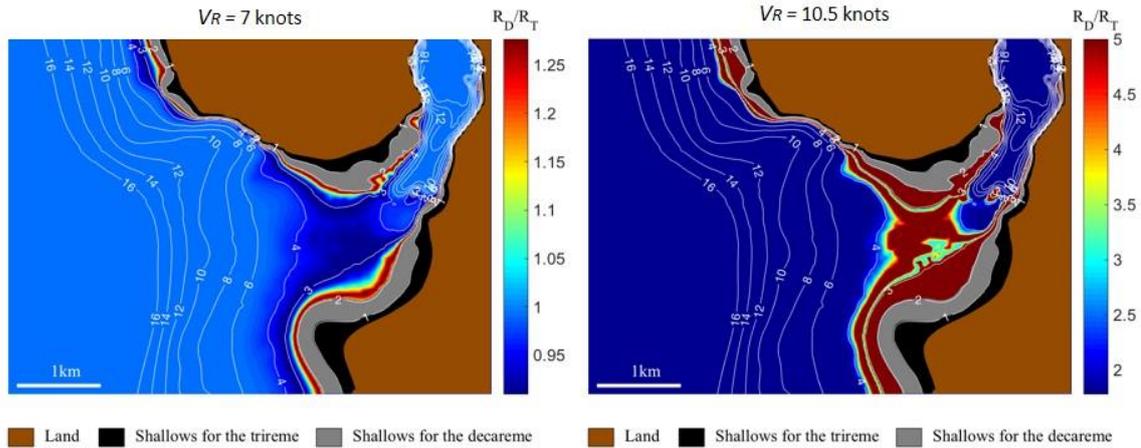

**Figure SI9.** Maps featuring the ancient bathymetry and theoretical predictions of wave making resistances for two different velocities: 7 knots (left) and 10.5 knots (right, which is saturated). At 7 knots, the decareme's wave resistance $R_D$ can be lower than the trireme's wave resistance.

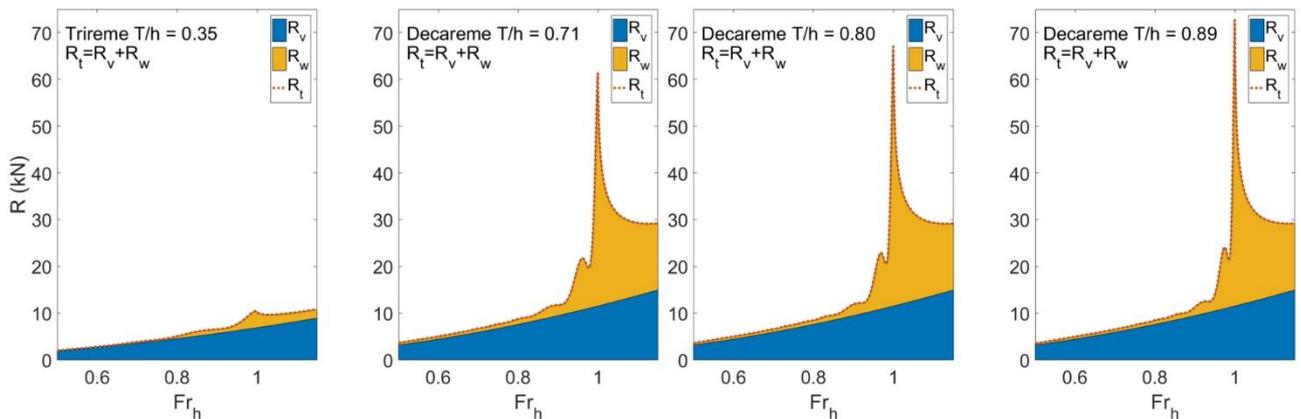

**Figure SI10.** Plots of total resistances broken down into viscous resistance (blue) and wave making resistance (yellow), for trireme and decareme configurations. The viscous resistance is calculated with the method (ITTC, 1957), and the wave making resistance with the Sretensky's formula.

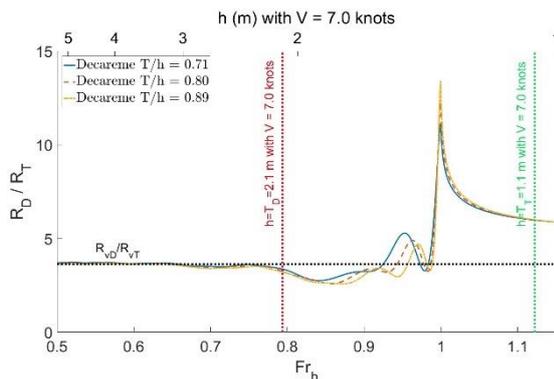

**Figure SI11.** Ratio between total resistances of the decareme and of the trireme, with a fixed velocity of 7.0 knots. The ratio of total resistances tends towards the ratio of viscosity resistances, equal to 3.6 for low Froude numbers ($Fr_h < 0.75$ and $h > 2.5$ m). The peak of resistance is not undergone by the decareme since it is in a zone of too shallow depth (the decareme's draft is bigger than the depth $T_D > h$).

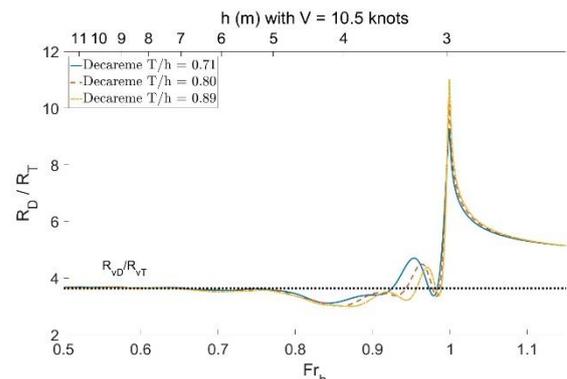

**Figure SI12.** Ratio between total resistances of the decareme and of the trireme, with a fixed velocity of 10.5 knots. The ratio of total resistances tends towards the ratio of viscosity resistances, equal to 3.6 for low Froude numbers ($Fr_h < 0.75$ and $h > 5$ m).

The use of the ITTC57 protocol is consistent with the experiments conducted. Indeed, if we compare this method to that used by (Coates, 1989), based on the towing tank measurements of (Grekoussis and Loukakis,

1986), we find similar results (Figure SI13). The computed viscous resistance with IITC57 is close to Coates' viscous resistance, up to 8 knots (superior to the cruising speed of 7 knots), the speed for which the wave resistance starts to play a role in deep water. According to Coates' curves (Figure SI13; Coates, 1989), from cruising speed (7-8 knots) the wave resistance starts to play a role. Thus, a greater effort must be made to increase the speed of the ship. Even in deep water, the wake clings to the ship as the echeneis-remora, and increases the difficulty in reaching the attack speed. Shallow water effects can totally prevent reaching this speed by a "wall of resistance" (Figure SI9).

In the future we may have to switch to a nonlinear Rankine-source panel method.

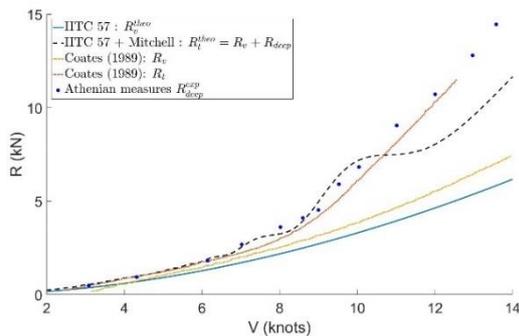

**Figure SI13. Comparisons between the experimental results of (Grekoussis and Loukakis, 1986), the ITTC57 calculation with Michell's theory, and the results of (Coates, 1989). Michell's theory is the limit of Sretensky's formula for $h$ infinite (Kirsh, 1966).**

## 9.6 SUPPLEMENTARY INFORMATION ON FLUID MECHANICS

### 9.6 (a) Experimental Materials and method of stereovision

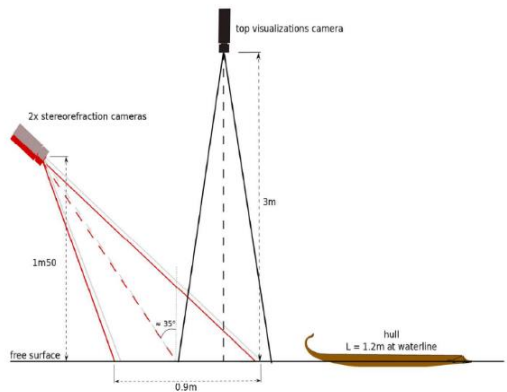

**Figure SI14. Sketch of the experimental setup with the positions of the recording cameras. The stereovision method is based on the deformations of a random dots pattern glued to the bottom.**

The experiments have been carried out in the towing tank of the Pprime Institute in Poitiers, France. Its geometry is a rectangular section with 1.5 m wide and a water level up to 1.2 m. The channel is 20 m long, and the measured zone, where we placed cameras, was at 10 m from the starting. The ship is towed by a trolley along the longitudinal axis x of the channel, with a speed up to 2.35 m/s. Thus, the maximum velocity of the decareme is 33 knots, and the trireme's maximum velocity is 23 knots, at real-scale. The acceleration had been fixed at 0.5 m/s², and a computer controlled the trolley's velocity. The ship was fixed (no translation or rotation), in order to test the effect of the Antonian number $A_n=T/h$ only.

As the amplitudes of the waves produced by a hull with such a length are very small, the method to measure these waves needs to be very precise and with a high resolution since the often used methods such as intrusive resistive probes cannot be applied. Such an optical method would have been difficult to apply with a bigger reduced model (such as the one of Grekoussis and Loukakis, 1985; 1986) since the extent of the visualizations windows would increase as well as the size of the data post-processing. Moreover, the limited resolution of the cameras would have been a restriction for such a large field of investigation. The wake-patterns have been measured with an original stereovision method, inspired from an earlier method developed in our team (Chatellier et al., 2013; Gomit, 2013b; Caplier, 2015) and improved for our purpose (see below). Two SpeedSense 1040 cameras from Dantec Dynamics with 28mm focal lenses have been placed above the water surface with a relative angle (Figure SI14) to capture the deformations of the random dots pattern, a roughcast of 750mm x 200mm (half the width of the water channel). The first step of the measurement is to record the image of the pattern with free surface at rest (Figure SI15). Then, the boat is launched and the two cameras capture images of the pattern deformed by the free surface undulations caused by the passage of the ship, at a frequency of 10 frames per second (fps) during 20s. The second step is to calculate the displacement of the water level in pixels on each image of the cameras, and then to calculate the displacement in millimeter for each time step. At the end we can reconstruct the whole wake behind the ship. The synchronization of the cameras and the acquisition of the images are performed with the DynamicStudio software. Each trial is recorded 3 times and then the images are processed and data are correlated with a dedicated algorithm. Then we can reconstruct the wake with a vertical precision of 0.1 mm on the water height and a horizontal spatial resolution of 10 mm (Figure SI17). The stereorefraction method has been validated with measurements of the wake produced by a 1.2 m long Wigley hull with rectangular geometry (the archetype of laboratory ships) for the deep-water configuration in the towing tank. The wake produced by this reference boat is the usual Kelvin wake in deep water (Figure SI16). Visualizations of the wake have been made from the top with a Jai CV-M2 camera with a 14mm focal lens at a frequency of 20fps.

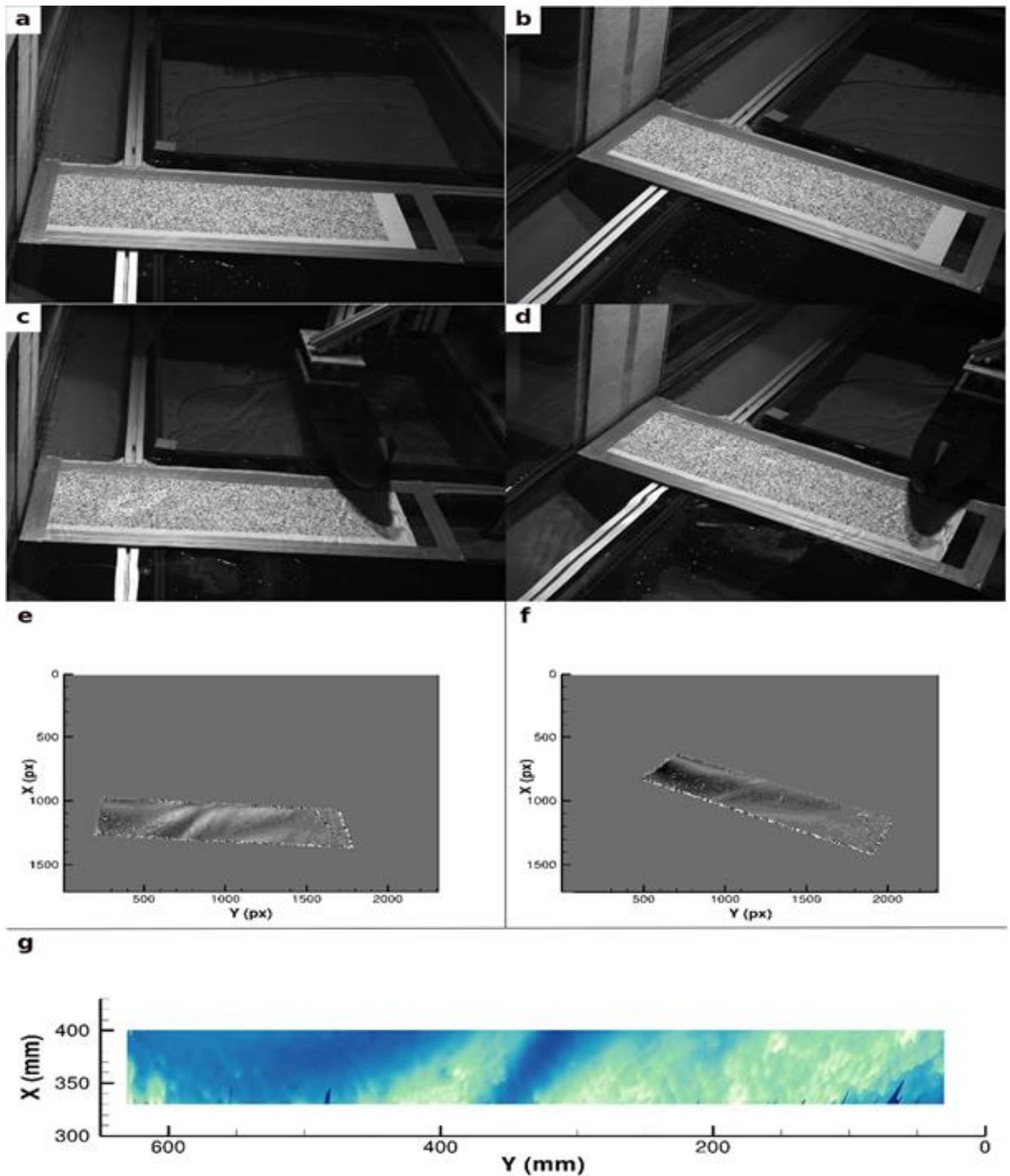

**Figure SI15.** The steps of the stereorefraction method from the acquisition of the reference images to the reconstruction of the surface deformation. Each column corresponds to one camera. a-b, Images of the reference pattern with the free surface at rest. c-d, Images of the pattern deformed by the free surface undulations. e-f, Free surface deformation in pixels (calculated with a correlation algorithm). g, the free surface deformation in millimeters (calculated with a reconstruction algorithm).

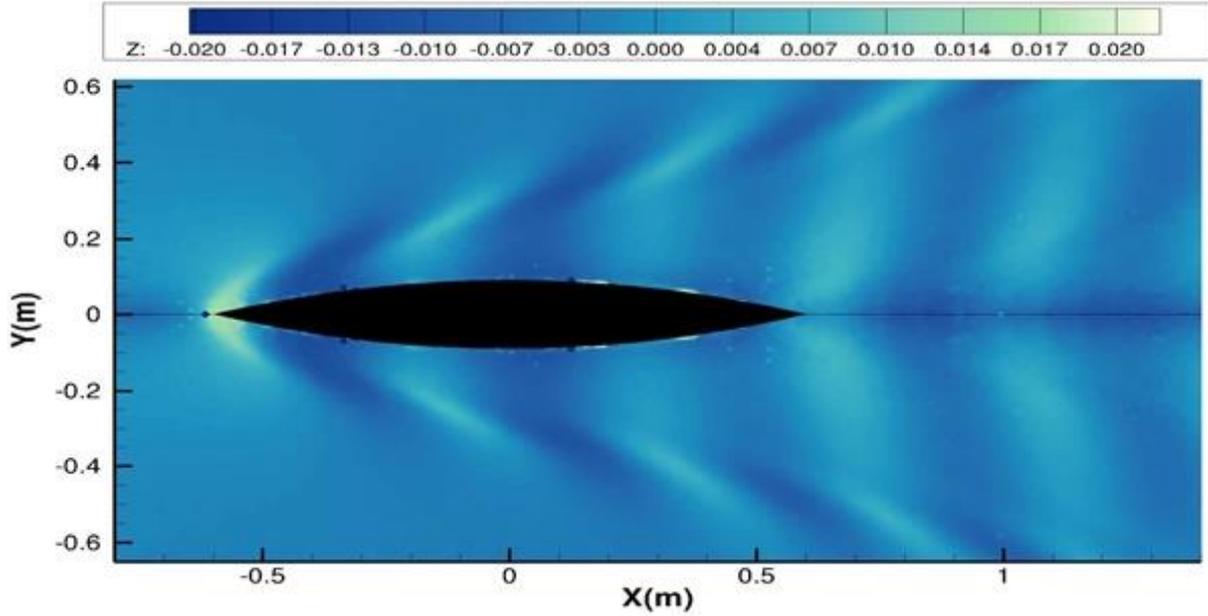

**Figure SI16.** The usual Kelvin deep water wake produced by a Wigley hull, measured with the stereorefraction method in the towing tank.

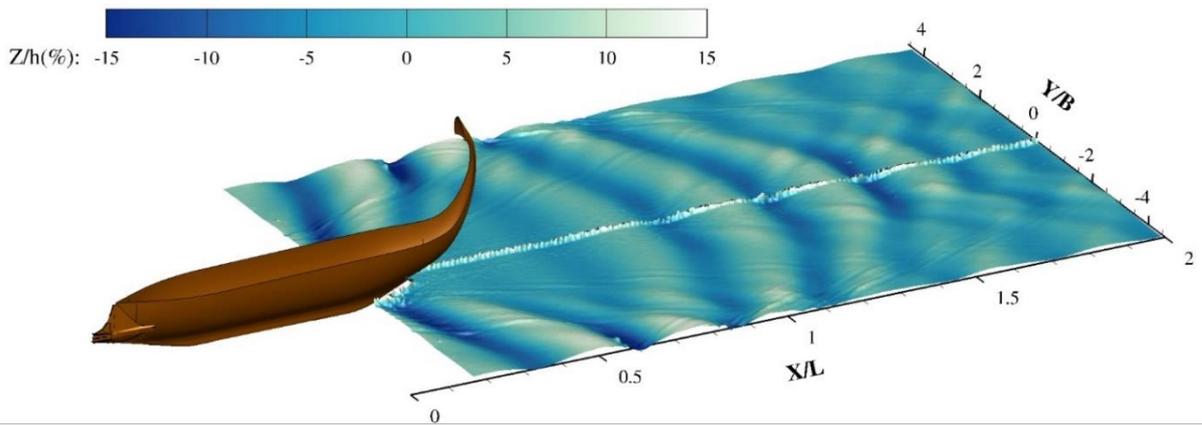

**Figure SI17.** The peculiar echeneidian wake behind the ship in the decareme configuration at $Fr_h=0.9$, measured with the stereorefraction method in the towing tank. The boat has no angle with the horizontal.

9.6 (b)  Use of spectral domain

Extended views of the wake allow spectral analysis of the wake, according to the method presented (Carusotto and Rousseaux, 2013), used in deep water by (Gomit, 2014), or in confined configuration (Caplier, 2015). By selecting the stern wake, in the real space (x, y), and with a Discrete Fourier Transform, we get a representation in the Fourier space ($k_x$, $k_y$). The spectral domain brings a lot of additional information to the visualizations: the energy distribution in the wake spread over different wave numbers, or the hydraulic response around the ship. Energy is distributed along the dispersion relation:

$$0 = V_m^2 k_x^2 - \left(gk + \frac{\sigma}{\rho}k^3\right)\tanh(kh)$$

We can bring out several remarkable values: the cutoff wave number $k_c$, the intersection between the dispersion relation and the abscissa axis; the inflection point $k_x^{infl}$, where the slope is a measure of the angle of the wake: $\tan(\alpha) = \left(\frac{dk_y}{dk_x}\Big|_{infl}\right)^{-1}$. Wave numbers before and after the inflection point are respectively relative to the transverse and divergent waves. The hydrodynamic response, which depends on the speed and shape of the boat, feeds the lower part of the spectral domain with $k_x < k_c$ (Carusotto and Rousseaux, 2013; Gomit, 2013a, 2014) (Figure SI18).

The Figure SI19 shows the selection of a part of the wake, its representation in the spectral domain, and the detection of the dispersion relation. After a detection of amplitude maximums, a polynomial interpolation is carried out on the experimental measurements, in order to measure a slope and the angle of wake. Bounded harmonics are observed at high speeds hence the appearance of additional branches at high wavenumbers with a corresponding non-linear deformation of the wake in the real space.

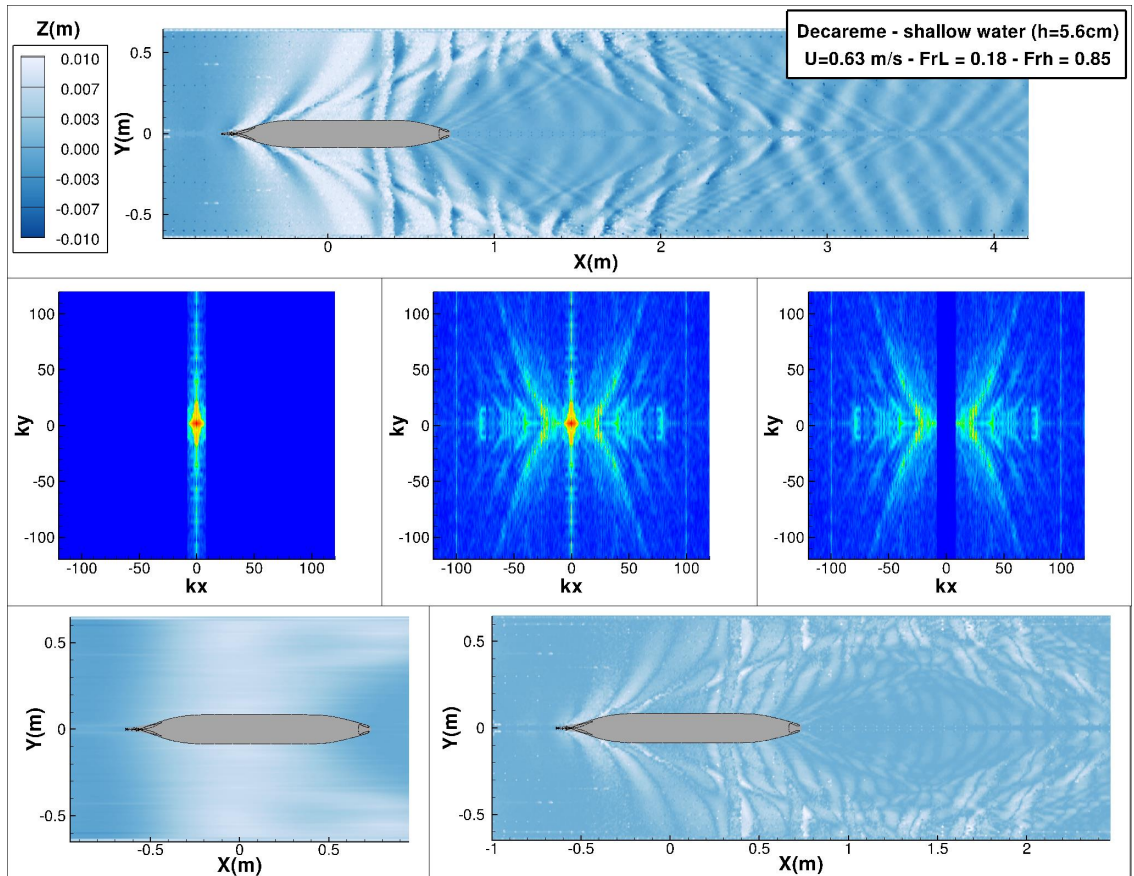

**Figure SI18. Fourier space of a wake measured by stereo-refraction and effect of the filter: hydrodynamic response ($k_x < k_c$), wake ($k_x > k_c$). The keel of the boat makes an angle of 0.13 ° with the horizontal (stern sunk).**

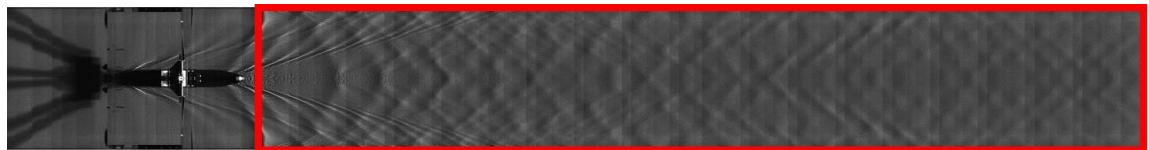

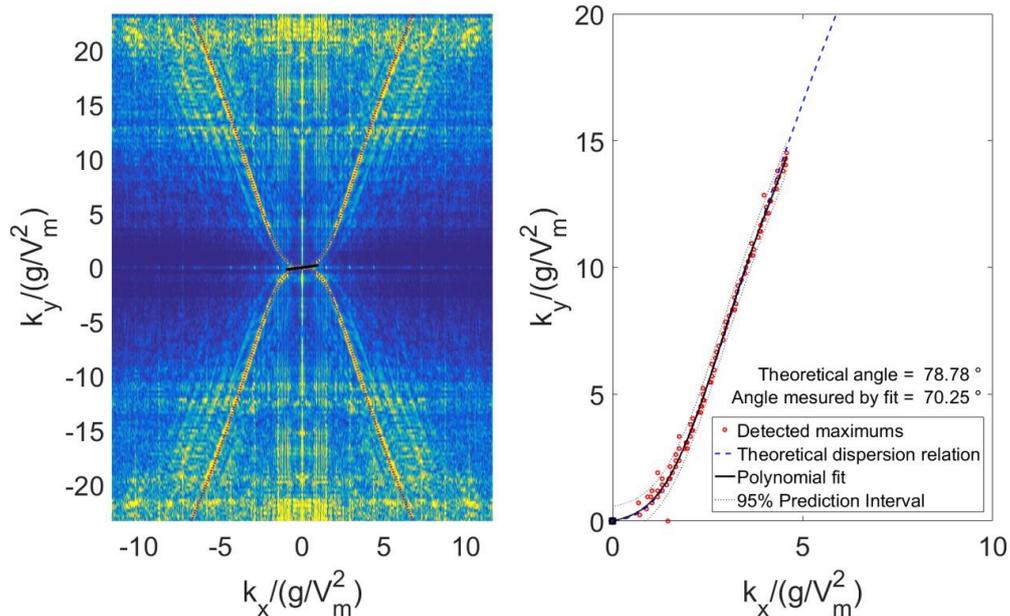

**Figure SI19. Processing of a top view of the echeneidian wake by Fast Fourier Transform at $Fr_h$=1.02. The FFT is done on the red box with a simple visualization by a top camera.**

### 9.6 (d) reduced model limits

On a real scale, the capillary term $k^3$ in the dispersion relation can be neglected, and the waves considered as purely gravity. However, by carrying out the reduced model experiments, this term can have an impact. Thus, by calculating theoretically the cut-off wavenumber for the transverse waves with ($k_c$) and without surface tension ($k_c^\sigma$), we may find a difference of up to more than 100% (Figure SI20). In the case of our experiments, the difference does not exceed 4%, for the lowest speeds tested in decareme configuration. Thus, the scale of our model does not involve significant additional capillary effects

$$k_y = 0 \Rightarrow k = k_x = k_c$$
$$0 = V_m^2 k_c^{\sigma 2} - \left(g k_c^\sigma + \frac{\sigma}{\rho} k_c^{\sigma 3}\right) \tanh(k_c^\sigma h)$$
$$0 = V_m^2 k_c^2 - g k_c \tanh(k_c h)$$

The reduced model tests are also limited by the velocity $V_m = 23$ cm/s (Rousseaux et al., 2010), below which the wake is killed by capillary effects (black area on the Figure SI20).

The experiments are carried out in a towing tank, which generates additional confinements to the vertical confinement ($A_n = T/h$): a transverse confinement ($B/W$), and a "sectional confinement" ($m=A_b/A_c$). The first leads to reflections of the wake on the walls of the canal (reflections creating interferences). The second generates a return current and causes a water level drawdown of the ship (Pompée, 2015). According to Schijf's theory, some of tested configurations are in transcritical regime. With side visualizations, we checked that the model did not lead to a significant water level drawdown of the ship. The transcritical effect becomes important only for excessive speeds ($V_R > 12$ knots).

Because we have chosen a geometric scaling and a Froude scaling, hence the Reynolds scaling cannot be respected. Thus, we have a factor between the real-scale Reynolds number, and the Reynolds number model:

$$\text{Re}_m = V_m L_m \frac{1}{\nu} = \frac{V_R}{\sqrt{\lambda}} \frac{L_R}{\lambda} \frac{1}{\nu} = \frac{\text{Re}_R}{\lambda^{3/2}}$$

In decareme configuration, we have $\text{Re}_m = \frac{\text{Re}_R}{392}$, and in trireme configuration, $\text{Re}_m = \frac{\text{Re}_R}{138}$. $\text{Re}_R$ is in the order of $10^8$, thus, $\text{Re}_m$ is in the order of $10^6$, so we are still in a turbulent Reynolds number regime.

As the Reynolds scaling is not respected, we have to calculate the viscosity in order to take into account the effect of scales. Doutreleau et al., (2011) give a calculation of the kinematic viscosity. For the model:
$$\nu = ([0.585(t_W - 12)10^{-3} - 0.03761](t_W - 12) + 1.235)10^{-6}$$
For the real-scale:
$$\nu = ([0.659(t_W - 1)10^{-3} - 0.05076](t_W - 12) + 1.7688)10^{-6}$$

With $t_W$ the water temperature ($t_W$ was 15°C at Actium, and 21°C in our towing tank).

Finally, since the channel has a finite length (20 m), and our measurement zone was 10m from the starting point, unsteady effects may occur. However, apart from the high speeds tested but unrealistic at full scale (exceeding 12 knots), the wake appeared to have stabilized as the boat passed through the study area (see Robbins et al., 2011; Macfarlane and Graham-Parker, 2018).

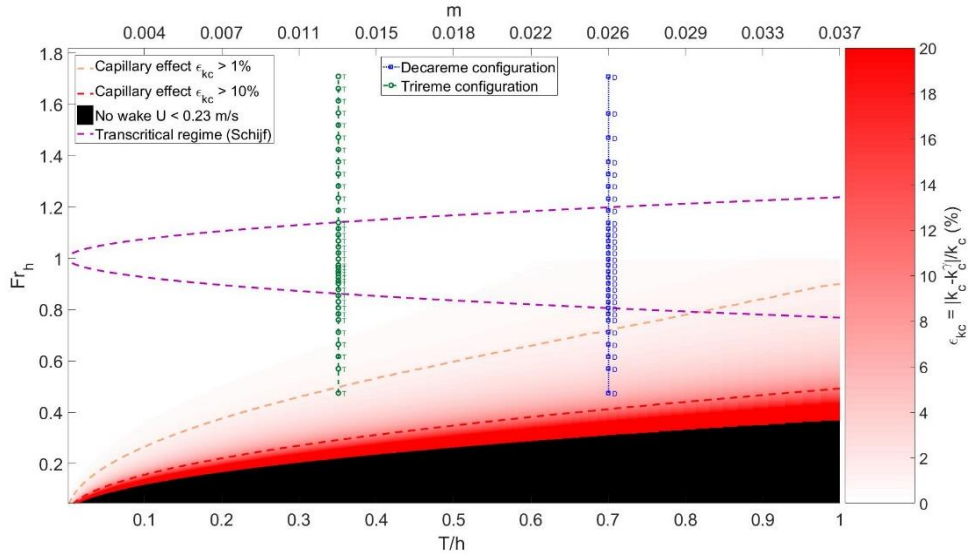

**Figure SI20. Capillary effect as a function of the ratio *T/h*, and the Froude. In the black zone, the capillary effects suppress the wake. Green *T* and blue *D* are configurations made in the towing tank. Purple lines are the borders of the transcritical regime according to Schijf's theory.**